\documentclass[10pt]{iopart}

\expandafter\let\csname equation*\endcsname\relax
\expandafter\let\csname endequation*\endcsname\relax
\usepackage{graphicx,amsmath,comment,color}
\usepackage{hyperref}

\begin{document}

\title{Hydrodynamic instabilities in miscible fluids}

\author{Domenico Truzzolillo and Luca Cipelletti}

\address{Laboratoire Charles Coulomb (L2C), UMR 5221 CNRS-Universit\'{e} de Montpellier,
4 F-34095 Montpellier, France}
\ead{domenico.truzzolillo@umontpellier.fr}
\vspace{10pt}
\begin{indented}
\item[September 29th 2017]
\end{indented}

\begin{abstract}
Hydrodynamic instabilities in miscible fluids are ubiquitous, from natural phenomena up to geological scales, to industrial
and technological applications, where they represent the only way to control and promote mixing at low Reynolds numbers, well below
the transition from laminar to turbulent flow.
As for immiscible fluids, the onset of hydrodynamic instabilities in miscible fluids is directly related to the physics of their interfaces.
The focus of this review is therefore on the general mechanisms driving the growth of disturbances at the boundary between miscible fluids,
under a variety of forcing conditions. In the absence of a regularizing mechanism, these disturbances would grow indefinitely.
For immiscible fluids, interfacial tension provides such a regularizing mechanism, because of the energy cost associated to the creation
of new interface by a growing disturbance. For miscible fluids, however, the very existence of interfacial stresses that mimic an effective
surface tension is debated. Other mechanisms, however, may also be relevant, such as viscous dissipation.
We shall review the stabilizing mechanisms that control the most common hydrodynamic instabilities, highlighting those cases for which
the lack of an effective interfacial tension poses deep conceptual problems in the mathematical formulation of a linear stability analysis.
Finally, we provide a short overview on the ongoing research on the effective, out of equilibrium interfacial tension between miscible fluids.
\end{abstract}
\tableofcontents
%\maketitle
\ioptwocol

\section{Introduction}
Hydrodynamic instabilities occupy a special position in fluid mechanics, being of fundamental importance for the deep understanding of a
variety of natural phenomena and for many applicative purposes.
Whether the boundary between miscible or immiscible fluids is stable to a random disturbance impacts, e.g., the dendritic shape of snow
flakes \cite{langer_instabilities_1980}, early failure in zinc alkaline batteries \cite{gallaway_lateral_2010}, breath sounds caused by surfactant
deficiency in lungs \cite{huh_acoustically_2007} and the protection of the stomach from its own gastric acids \cite{bhaskar_viscous_1992}.
Morphological patterns and periodic structures are often the result of interfacial instabilities. They can be beneficial, like in the case
of chromatographic separation where viscous fingering can improve mixing in non-turbulent systems and small-scale devices \cite{jha_fluid_2011},
or disadvantageous, like for oil recovery and pipe cleaning, where the interfaces between the cleaning and the waste fluid must stay stable \cite{Lake1989} to optimize the efficiency of the removal process.
Hence, depending on the application, either a stable or an unstable interface could be desirable,
making the ability to control interfacial instabilities essential in technological applications.

The theory of hydrodynamic instability forms a substantial part of the arsenal of techniques available to the researchers in fluid mechanics for
studying and controlling stable-to-unstable flow transitions in a wide variety of flows in physics, mechanical and chemical
engineering, aerodynamics, and natural phenomena (climatology, meteorology and geophysics).
The literature on this subject is so vast that very few researchers have attempted to write a pedagogical text which describes the major developments in the field.
Classical instability theory essentially deals with flows in porous media and quasi-parallel or parallel shear flows such as mixing layers,
jets, wakes, Poiseuille flow and boundary-layer flow. Such configurations are the focus of the books by Drazin and Reid \cite{Drazin2004},
by Schmid and Henningson \cite{Schmid2001}, and by F. Charru \cite{charru_hydrodynamic_2011}. They are of particular interest to all those
researchers who investigate stress-induced deformations.

The stability problem of miscible two-fluid systems is of particular importance because miscible interfaces are ubiquitous in nature and industry.
Sharp gradients of concentration, temperature, density or viscosity define boundaries between miscible fluids: this is the case of
boundaries defining ocean currents \cite{neumann_ocean_2014} and silicate fluids in earth mantle \cite{Morra2008}, or the spontaneous formation of
a cell-free layer in blood microcirculation \cite{fedosov_blood_2010}. Turbulent flows at high Reynolds numbers are effective mixers because
of the chaotic nature of the velocity field and of the energy cascade spanning a wide range of length scales \cite{Batchelor1959}. By contrast,
at low Reynolds numbers, inertial effects are negligible, turbulence does not set in and hydrodynamic instabilities represent
a unique strategy to enhance mixing, which motivates the importance of understanding stability in miscible fluids.

\begin{figure*}[htbp]
  \centering{\includegraphics[width=15cm]{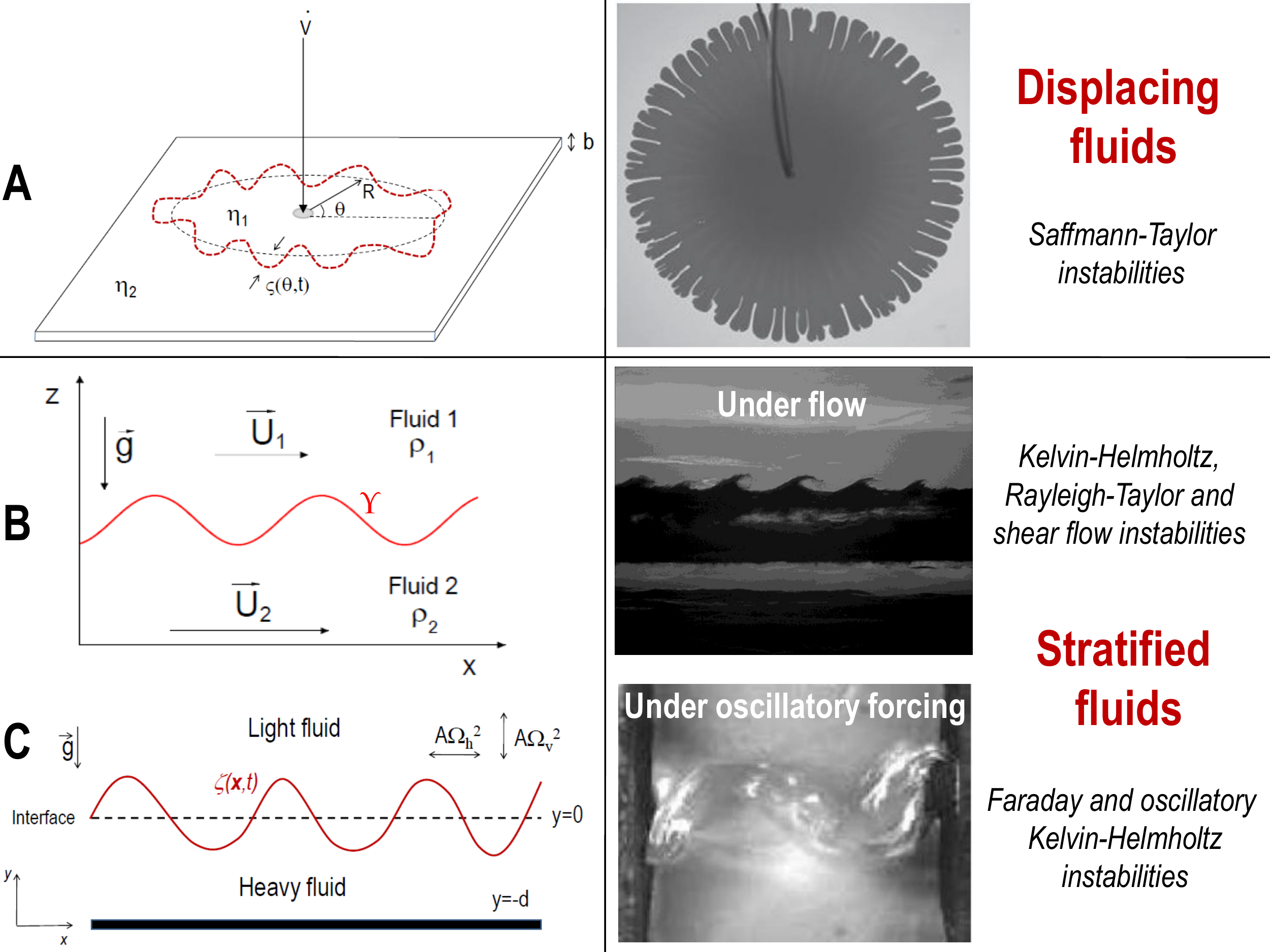}}\\
  \caption{
  Overview of the fluid dynamic instabilities occurring in the presence of miscible interfaces.
  A, left: Schematic configuration of radial source flow. The viscosities of the displacing and displaced fluids are $\eta_1$ and $\eta_2$,
  respectively. Fluid 1 is injected into a Hele-Shaw cell, previously filled with fluid 2, at a constant injection rate $\dot{V}$. $b$ is the gap
  between the plates of the cell. The dashed black line represents the time-dependent average position of the interface $R(t)$ and the red
  undulated curve depicts the perturbed interface $\Upsilon(\theta,t)=R(t)+\varsigma(\theta,t)$, where $\theta$ is the polar angle. Right: a
  viscous fingering pattern observed by injecting a water-glycerin mixture in a cell filled with another (more viscous) water-glycerin mixture with
  $\eta_2=8.13\eta_1$ (adapted with permission from \cite{bischofberger_fingering_2014}).
  B, left: Sketch of unstable two-fluid coflow with velocity discontinuity. The interface $\Upsilon$ separates
  two fluids of different density $\rho_1$, $\rho_2$ flowing with different mean laminar velocities $U_1$, $U_2$. Right: Kelvin-Helmholtz
  instability observed in clouds over San Francisco. These clouds, sometimes called "billow clouds," are produced when horizontal
  layers of air having different densities brush by one another at different velocities (adapted from \cite{wikiKH},
  licensed under the Creative Commons Attribution-Share Alike 4.0 International). Panel C: Sketch of an unstable interface between two fluids of different density under oscillatory forcing.
  Faraday and Kelvin-Helmhotz-type instabilities may occur under vertical and horizontal forcing characterized by fixed angular frequencies $\Omega_v$ and $\Omega_h$ respectively. On the right a 1D-Faraday instability observed at the interface between pure and salt-saturated water
  (adapted with permission from \cite{zoueshtiagh_experimental_2009}).}
  \label{summary}
\end{figure*}

This review surveys the general mechanisms driving hydrodynamic instabilities in miscible fluids,
giving emphasis to the consequences arising from the lack of a true equilibrium interface, since fluids intermix over time.
We will draw analogies and point out differences between miscible and immiscible interfaces, discuss the theoretical problems arising
from fluid mixing and argue the presence of subtle stabilizing phenomena, whose presence represents an important and controversial issue.
%%%%The instabilities occurring in presence of miscible interface are the following%%%%
We will discuss instabilities occurring in the presence of a preexisting (unperturbed) boundary between two homogeneous, miscible fluids.
These are summarized in Figure \ref{summary}, where instabilities are divided into two distinct families: i) those occurring when a driven less
viscous fluid displaces a more viscous one (Saffman-Taylor instabilities) and ii) those arising in a stratified configuration and caused
by tangential motion, a gravitationally-unstable density stratification, a viscosity stratification or
an oscillatory forcing (Kelvin-Helmholtz, Rayleigh-Taylor, shear
flow, Faraday and oscillatory Kelvin-Helmoltz instabilities, respectively).\\
%Parlare di quali sono stabilizzate e quali no da tensione/viscosità e quali no
Whether or not all these instabilities appear depends on the balance between the mechanisms, specific to each instability, that enhance the
amplitude of a disturbance and those tending to stabilize the interface. Among the latter, capillary forces often play a primary
role. The assumption of zero surface tension, often invoked for miscible interfaces, leads therefore to an ultraviolet catastrophe, making the
problem ill-posed: the shorter the wavelength of the disturbance, the faster the growth rate of the unstable mode associated to it.
This is the case of inviscid and viscous Kelvin-Helmholtz instability, and of the Saffman-Taylor and Rayleigh-Taylor instabilities.
Understanding in which cases capillary forces are at work in miscible fluids is therefore an important aspect of the stability
problem. For such a reason, it will be discussed in the second part of this review in some depth.

The review is organized as follows: in section \ref{Hadamard} we introduce the aforementioned
pathology affecting some instabilities when surface tension is zero: the so-called Hadamard instability. This will help the reader
to understand the immiscible-to-miscible crossover and the conditions under which
the stability problem becomes catastrophic. A brief summary of the more recent attempts to regularize it will be presented.
We will then discuss the causes of
disturbance growth at miscible interfaces in different configurations and forcing conditions. In section \ref{SF} we will analyze the case of
displacing fluids and the onset of viscous fingering. In section \ref{KHRT} we will give an overview of the instabilities arising in stratified
fluids at rest and under flow when a stationary velocity mismatch between the two fluids exists. This is the case of the Kelvin-Helmholtz and
Rayleigh-Taylor waves. In sections \ref{Faraday-sec} and \ref{KHOsc} we will debate the stability of miscible interfaces under vertical and
horizontal oscillatory forcing, respectively.
Finally, we dedicate the last part of this review to the quite debated issue concerning the existence of capillary forces at the boundaries
between miscible fluids. We emphasize that we consider this issue as a crucial one for the correct mathematical formulation of
the linear stability problem in miscible interfaces and the interpretation of the experiments.
We will recall the main experimental evidences demonstrating the existence
of true interfacial stresses, and the attempts to rationalize their existence.

%Finally we propose a simple though significant model on lattice that allows to calculate an effective tension between miscible fluids.

\section{Hydrodynamic instabilities in miscible fluids: well-posedness and driving mechanisms}

\subsection{A dutiful premise: Hadamard instabilities and ill-posed problems}\label{Hadamard}
To determine the stability of a fluid flow one must consider how a fluid reacts to a disturbance. These disturbances, that can always be expressed
as a sum of sine waves, are usually considered as related to the initial physical properties of the fluid such as velocity, pressure and density.
The so-called linear stability analysis \cite{drazin_hydrodynamic_2004} of the flow defines accurately the concept of 'stability' or 'instability'
based on the fluid response to an infinitely small disturbance containing all wavelengths, each component evolving independently.
A system will be stable if,
at any time, any infinitely small variation, which is considered a disturbance, will not have any noticeable effect on the final state of the
system and will eventually die down with time. We say, in this case, that the disturbance does not evolve with a strictly positive growth rate.
One of the most important, and in some cases the sole, stabilizing effect at the interface between two fluids is surface tension, that suppresses
the growth of small-wavelength perturbations. For this reason, the analysis of miscible interfaces, where equilibrium surface tension does not
exist, gives rise to some conceptual problems, unless transient capillary forces are taken into account. D.D. Joseph has masterfully discussed
the problem \cite{joseph_fluid_1990, joseph_short-wave_1990} of short-wave instabilities with a growth rate that increases without bound as the
wavelength tends to zero, in the absence of interfacial tension. Such cases represent a so-called ill-posed problem: the models do not capture the
real physical phenomenon and catastrophically diverge in numerical analysis. Although nineteenth-century mathematicians contributed to the early
study of ill-posed problems, it is generally agreed that the subject came
to prominence only after Jacques Hadamard had formulated his well-known definition \cite{Hadamard1902,Hadamard1922}.
At the beginning of the last century, unfortunately, he developed an adverse view of the subject which, on becoming widely accepted,
had the effect of inhibiting further study. His
objections were grounded in his celebrated counterexample of the Cauchy problem for Laplace's equation \cite{Hadamard1922}. In order for a global
solution to exist, Hadamard demonstrated that the initial conditions must satisfy a well-defined compatibility relation. Even in the unlikely
event that the relation is satisfied, he further showed that the solution in general does not depend continuously on the initial conditions.
Such behaviour convinced Hadamard that ill-posed problems lacked physical relevance and hence should be ignored.
%This became the prevailing attitude, and consequently, in partial differential equations at least, the research activity became confined to the standard initial boundary value problems.
It was only the growing insistence for a precise theoretical understanding in applied sciences, principally geophysics and fluid dynamics,
that renewed the scientific interest in the lack of mathematical stability.
According to Hadamard, a problem is well-posed (or correctly-set) if
\begin{enumerate}
  \item it has a solution
  \item the solution is unique
  \item the solution depends continuously on the initial conditions.
\end{enumerate}
The meaning of (i) is clear. (ii) When we call a solution "unique", we mean sometimes
unique within a certain class of functions. For example, a problem might have several
solutions, only one of which is bounded. In this case we say that the solution is unique in
the space of bounded functions. (iii) A solution depends continuously on data and
parameters if small changes in initial or boundary functions and in parameter values
result in small changes in the solution.
Obviously, the fulfillment of conditions (i) to (iii) is of fundamental importance in physics.\\
Many physical problems are modeled by the solution of a particular partial differential equation: Laplace, Poisson, Schrodinger and diffusion
equations are just a few examples of partial differential equations that describe phenomena in all fields of physics, including fluid dynamics.
Though the classical theory of partial differential equations deals almost completely with well-posed problems, ill-posed problems can be
scientifically interesting and rise challenging questions, like in the case of miscible fluids.
As an example, we recall here the emblematic discussion by Hadamard \cite{hadamard_lectures_1926,kabanikhin_definitions_2008} of the Laplace equation as an ill-posed problem.
Let us consider the Laplace equation

\begin{equation}\label{Laplace}
    \frac{\partial^2 \psi}{\partial x^2}+\frac{\partial^2 \psi}{\partial y^2}=0
\end{equation}
where the scalar field $\psi(x,y)$ is defined only in the half space
\begin{equation}\label{h-space}
    \wp=\left\{x,y;y>0;-\infty<x<\infty\right\}
\end{equation}
with boundary conditions
\begin{eqnarray}\label{init-values}
\psi(x,0)=0 \\
\frac{\partial \psi}{\partial y}(x,0)=\frac{1}{k^{\alpha}}\sin(kx),\quad \alpha>0
\end{eqnarray}
With this choice, the boundary conditions are bounded and tend to zero for small wavelengths $(k\rightarrow\infty)$.
The solution of this problem is \cite{hadamard_lectures_1926,kabanikhin_definitions_2008}
\begin{equation}\label{sol1}
    \psi(x,y)=\frac{1}{k^{1+\alpha}}\sin kx\sinh ky.
\end{equation}
Small initial values  of $\frac{\partial \psi}{\partial y}(x,0)$ lead to a solution characterized by unbounded oscillations for any
arbitrary small $y>0$ as the wavelength $\lambda=2\pi/k$ tends to zero, giving rise to a so-called \emph{ultraviolet catastrophe}. This lack of
continuity of the solution with respect to the boundary values represents the core of \emph{Hadamard instability} that affects many situations
of physical interest, spanning from fluid dynamics and rheology \cite{bressan_nonlinear_2011} to astrophysics \cite{Alexeev2017xi}.
In all these cases the well-posedness of the problem depends on the existence of some quantity that stabilizes short-wavelength disturbances.
A similar problem is encountered in some hydrodynamic instabilities in miscible fluids, where the growth rate is unbounded in absence
of interfacial stresses and/or viscous dissipation. This is indeed the case of Saffman-Taylor, Rayleigh-Taylor and Kelvin-Helmholtz
instabilities that will be discussed in the next sections.

We recapitulate here a few general yet crucial remarks concerning ill-posedness:

\begin{itemize}
\item From the point of view of a physicist a well-posed problem is such that realistic initial/boundary conditions yield physically
reasonable solutions.
\item Ill-posed problems cannot be simulated, because of the rapid and catastrophic increase of the initial disturbance and
due to the sensitivity to numerical noise.
\item Regularizing mechanisms are preferentially found in some neglected physical quantities that, however small they may be, stabilize
short wave lengths. This is, e.g., the case of surface tension and viscosity in fluid dynamic instabilities.
\end{itemize}

In what follows, we take a closer look to some of the most important hydrodynamic instabilities occurring in miscible fluids.
In some cases (Saffman-Taylor, Faraday and oscillatory Kelvin-Helmoltz instability), we will attack the stability problem in the presence of surface
tension and look at the zero-surface-tension limit, while in other cases (Kelvin-Helmholtz, Rayleigh-Taylor) we will assume
no capillary forces since the beginning. In all cases, we will briefly discuss the ill-posedness of the problem
and the proposed solutions to unphysical catastrophic behavior.

\subsection{Displacing fluids: Saffman-Taylor instabilities}\label{SF}
Viscous fingering, known as Saffman-Taylor instability\cite{Saffman1958}, is a hydrodynamic instability that occurs
when a lower viscosity fluid displaces a higher viscosity one in porous media or in Hele-Shaw flows,
the latter being defined as Stokes flows between two parallel flat plates separated by a small gap. %Fundamental facets of the displacement process, such as mixing efficiency, depend strongly on the type of pattern created by the uneven boundary of the fluids. The interface created by these fingering patterns affects mixing and therefore understanding how these patterns evolve is crucial in applications such as oil recovery and groundwater remediation \cite{Davis1988,Petitjeans1996b}.
Saffman-Taylor instabilities represent a very important topic for the description of the dynamical behavior of interfaces
in multi-fluid systems \cite{Davis1988,Petitjeans1996b} and are, as we will see, "Hadamard-unstable" in the limit of vanishing
surface tension \cite{Joseph1990}.\\

The first scientific study of viscous fingering occurring between miscible fluids can be reasonably attributed to Hill (1952)
\cite{hill_channelling_1952}, who not only published a first simple "one-dimensional" stability analysis, but also conducted a series of
careful and quantitative experiments, recognizing and quantifying the destabilizing role of the viscosity mismatch between fluids.
He considered two distinct cases: 1) gravity-stabilized and pressure-driven viscous fingering in vertical down-flow of a lighter, less viscous
fluid displacing a heavier and more viscous one, and 2) the opposite case of viscous stabilization of a gravitationally unstable
configuration where a heavier, more viscous fluid flows downwards in a lighter, less viscous fluid.
Few years later, Chouke \emph{et al.} (1959)\cite{Chuoke1959} and Saffman \& Taylor (1958)\cite{Saffman1958} published their now-classical
papers on viscous fingering including the effect of surface tension. Both papers, submitted within six months of one another,
contain essentially identical linear-stability analyses of a one-dimensional displacement.
As reported by Saffman \cite{saffman_viscous_1986}, G. Taylor was the first in 1956 to realize that two-dimensional flow in a
porous medium can be modelled by flow in a Hele-Shaw apparatus consisting of two parallel plates separated by a small gap $b$. In this configuration,
the average two-dimensional velocity $\mathbf{v}$ of a viscous fluid is related to the fluid pressure $P$ as
\begin{equation}\label{Darcy}
\mathbf{v}=-\frac{b^2}{12\eta}\nabla P, \quad \nabla \cdot \mathbf{v}=0
\end{equation}
where $\eta$ is the viscosity of the fluid. Equation \ref{Darcy} is identical to the Darcy law describing the fluid motion in a porous
medium with permeability $b^2/12$.

Here, we will not go through the stability problem as first discussed by Saffman \& Taylor, who dealt with viscous fingering in a rectangular
channel. Instead, following Miranda \& Widom \cite{Miranda1998}, we focus on the instability occurring in a radial geometry, which includes the
problem in a rectangular geometry as a limiting case with zero interface curvature and infinite flow rate \cite{Miranda1998}.

We consider two immiscible, incompressible, viscous fluids. In a Hele-Shaw cell, fluids flow in a narrow gap of thickness $b$ between two
parallel plates (see Fig.\ref{summary}-A). Here $b$ is smaller than any other macroscopic length scale in the problem: the system is
considered to be effectively two-dimensional. The viscosities of the inner and outer fluids are $\eta_1$ and $\eta_2$, respectively. The flows
are assumed to be irrotational, except at the interface.
The fluid 1 is injected into fluid 2 through a hole of radius $r_0$ placed on the top plate of the cell, at a constant flow rate $\dot{V}$,
equal to the volume of fluid injected per unit time. %Between the two fluids there exists a surface tension $\Gamma$ that can be arbitrarily small and possibly zero.\\
During the flow the interface has a perturbed shape $\Upsilon(\theta,t)=R(t)+\varsigma(\theta,t)$ where $\theta$ represents the polar angle
and $R(t)$ the unperturbed radius of the interface,
\begin{equation}\label{radius-impert}
    R(t)=\sqrt{R_0^2+\frac{\dot{V}t}{b\pi}},
\end{equation}
where $R_0$ is the unperturbed radius at $t=0$. The evolution of the interface perturbation $\varsigma(\theta,t)$ can be described in terms of complex Fourier modes
\begin{equation}\label{Fourier-modes}
    \varsigma_n(t)=\int_0^{2\pi}\varsigma(\theta,t)e^{-in\theta} d\theta
\end{equation}
where $n$ is the mode number of the perturbation.\\

The evolution of the profile is obtained by assuming that the initial interface is affected by white noise (all $\varsigma(0)$ are equal)
and by calculating the evolution of the $\varsigma_n$'s.
By solving Eq. \ref{Darcy} in Fourier space and imposing a finite non-zero Laplace pressure between the two fluids,
Miranda and Widom \cite{Miranda1998} showed that $\varsigma_n$'s must fulfill a set of differential equations for $\varsigma_n/R\ll 1$ and
$n \neq 0$:
\begin{equation}\label{diff-eq-modes}
  \dot{\varsigma}_n=g(n)\varsigma_n \hspace{0.5cm} n=1,2...
\end{equation}
where
\begin{equation}\label{growthrate1}
  g(n)=\left[\frac{\dot{V}}{2\pi b R^2}(A|n|-1)-\frac{\alpha}{R^3}|n|(n^2-1)\right]
\end{equation}
represents the growth rate of the mode number $n$ in the linear regime. Here $$A=\frac{\eta_2-\eta_1}{\eta_1+\eta_2}$$
and $$\alpha=\frac{b^2\Gamma}{12(\eta_1+\eta_2)},$$ where $\Gamma$ is the interfacial tension.
From Eq. \ref{growthrate1} we get two important features of radial viscous fingering:
\begin{enumerate}
  \item the existence of a series of critical radii at which the mode $n$ becomes unstable because $g(n)$ switches from negative to positive;
  \item the presence of a fastest growing mode $n_f$, given by the closest integer to the maximum of $g(n)$.
\end{enumerate}
The critical radii $R_c(n)$ are those for which $g(n)=0$, i.e
\begin{equation}\label{critical_radius}
  R_c(n)=\frac{2\pi\alpha b}{\dot{V}}\frac{|n|(n^2-1)}{A|n|-1}.
\end{equation}
The maximum growth rate mode is determined by imposing $dg(n)/dn=0$ and reads
\begin{equation}\label{max growth}
  n_f=\sqrt{\frac{\dot{V}R(t)A}{2 \pi \alpha b}+1}.
\end{equation}
Note that $n_f$ is time dependent and must not be intended as the mode observed in experiments at the onset of the instability, where the mode
with maximum amplitude $n_A$ dominates \cite{Truzzolillo2014, Dias2013}.
This is indeed an important feature of radial viscous fingering for which we have shown recently \cite{Truzzolillo2014} that a very good agreement
between the mode number observed in experiments and that predicted by the theory is obtained only by considering the maximum amplitude
mode. For large wave numbers, one finds \cite{Truzzolillo2014} that $n_A$ is proportional to $n_f$: $n_A=0.422n_f$.

From Eqs. \ref{critical_radius} and \ref{max growth} we immediately realize that the problem becomes (Hadamard) ill-posed
as $\Gamma \rightarrow 0$, corresponding to $\alpha \rightarrow 0$ in equation \ref{max growth}: short wave length modes are not stabilized,
the critical radius is strictly zero for each mode and the wave number associated to the fastest growth rate mode tends to infinity.
This particular feature is not affected by the geometry of the problem: in both radial \cite{Miranda1998} and rectangular
\cite{miranda_weakly_1998} geometries short wavelengths are not stabilized without surface tension.\\
Many question arise in the framework of the Hadamard-unstable scenario: is there a dominant wavelength characterizing the onset of viscous fingering in miscible fluids?
If yes, which are the physical phenomena determining it? Does it depend on the viscosity contrast, injection rate or cell geometry?
The answer to these questions remains highly debated as we briefly discuss hereafter.

After the seminal work by Hill \cite{hill_channelling_1952}, a further attempt to experimentally investigate fingering between miscible fluids
has been reported by Wooding \cite{wooding_growth_1969}, who however did not discuss the linear stability problem that provides a prediction
for the expected wavelength at the onset of the instability.
Only some authors \cite{Saffman1958,Chuoke1959,White1976,Paterson1981,nittmann_fractal_1985, Miranda1998} have explored the initial growth of
fingers and derived expressions for the incipient fingers with maximum growth rate or maximum amplitude \cite{Truzzolillo2014}. In all cases but
one \cite{paterson_fingering_1985}, the introduced models predict that, as interfacial tension tends to zero, the finger wavelength tends to zero
possibly with diverging growth rates.

\begin{figure}[htbp]
  \centering{\includegraphics[width=9cm]{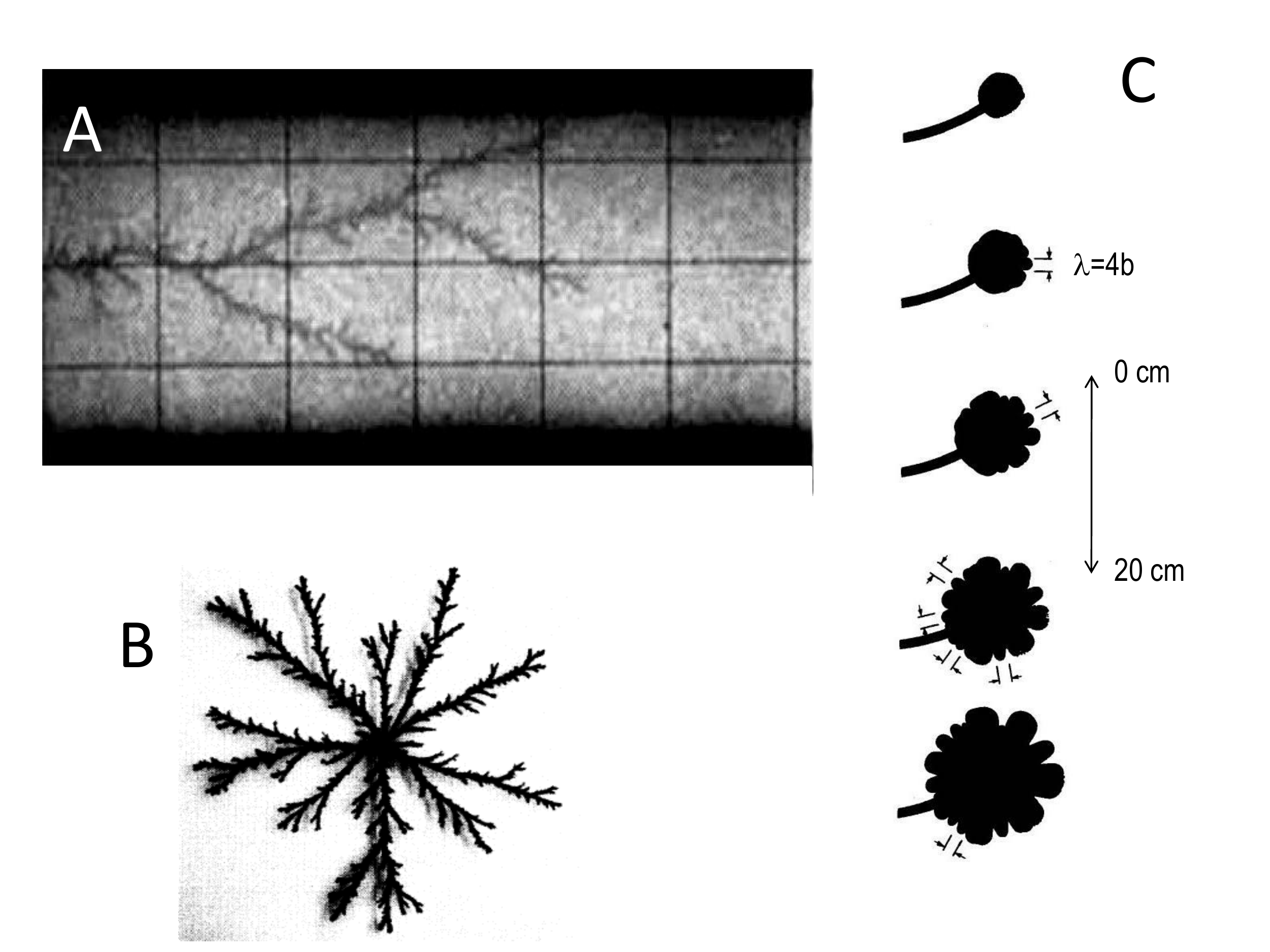}}\\
  \caption{Panel A: Viscous fingering created by water displacing a polysaccharide aqueous solution. The water is dyed with methylene blue and is
  injected by a single inlet in the left wall (adapted with permission from \cite{nittmann_fractal_1985} ).
  Panel B: Radial viscous fingering observed when water displaces a shear thinning polymer solution observed by \cite{daccord_radial_1986}.
  Panel C: Photographs of radial viscous fingering patterns observed when water is injected into a Hele-Shaw cell filled with glycerol.
  In this example $\dot{V}=0.255$ cm$^3$/s, $b=$ 0.30 cm. Photographs are taken at $t=39$, $65$, $94$, $125$ and $164$ s after the injection start.
  The distance $\lambda$ indicated by Paterson \cite{paterson_fingering_1985} is $4b$.}\label{Pat-Nit}
\end{figure}

Particular attention should be payed to the work of Nittmann \emph{et al.} \cite{nittmann_fractal_1985} and Paterson \cite{paterson_fingering_1985} since they reach important conclusions through quite different theoretical approaches. Nittmann and coworkers \cite{nittmann_fractal_1985}
argued that, in absence of surface tension, for an incompressible fluid of viscosity $\eta_1$ displacing a fluid of viscosity $\eta_2>\eta_1$
in a rectangular channel at mean velocity $v$, the velocity potential $\phi$, defined by the scalar field such that $v=-\nabla\phi$,
evolves according the Laplace equation $\nabla^2\phi=0$. The Laplace equation under fixed boundary condition,
i.e. when the potential $\phi$ in the less viscous fluid and at the channel outlet can be approximated by two constants,
is identical to the equation describing  diffusion-limited aggregation (DLA). In DLA
two particles bond to each other only if they collide by diffusion; particle aggregation leads to the formation of aggregates
with fractal morphology.
Based on this analogy Nittmann \emph{et al.} \cite{nittmann_fractal_1985} argued that, the expected late-stage fingering patterns of miscible fluids are
fractal-like. Such kind of patterns were indeed observed by the same authors for water displacing an aqueous solution of polysaccharides \cite{nittmann_fractal_1985}
in a rectangular geometry (Figure \ref{Pat-Nit}-A) and by Daccord \emph{et al.} \cite{daccord_radial_1986} for water displacing a highly viscous
aqueous polymer solution in the radial geometry (Figure \ref{Pat-Nit}-B). %Nevertheless fractal-like patterns have been observed by May and Maher \cite{may_fractal_1989} also for immiscible fluids, ruling out the possibility that such late-stage morphology is only due to the supposed absence of interfacial tension.
It's worth remarking that these results, strictly obtained for the late-stage evolution of the patterns under the assumption of infinite
capillary number (vanishing $\Gamma$), do not give a hint on a possible regularizing mechanism for the onset of viscous fingering.

Quite at the same time, Paterson \cite{paterson_fingering_1985} used perturbation theory to investigate the early-stage evolution
of radial viscous fingering. He showed that viscous dissipation increasingly damps fingers as the shear rate in the cell increases, and that the
problem should be considered as three-dimensional. By using the principle of minimum entropy production \cite{Prigogine1955}, Paterson showed
that the initial radial viscous fingering between miscible fluids should be characterized by a cut-off wavelength $\lambda_{c}=4b$ that does
not depend on the injection rate nor the viscosity contrast. He found good agreement between such prediction and the patterns observed by
injecting water in a cell filled with glycerin (Figure \ref{Pat-Nit}-C). The result of Paterson has been further reproduced more recently by
Bischofberger \emph{et al.} \cite{bischofberger_fingering_2014} who employed water-glycerol mixtures of different viscosities and compositions.
Collectively these experiments show that the theory of Paterson captures the behavior of molecular newtonian liquids where
interdiffusion rapidly smears the concentration profile, such that capillary forces can be neglected.
%Daccord et al. \cite{daccord_radial_1986} for water displacing non-newtonian polymer suspensions and, more recently
On the other hand, we have recently shown \cite{TruzzolilloPRX2016} that the wavelength characterizing the onset of radial viscous fingering,
does depend on the injection rate and, most importantly, on the physico-chemical nature of the two fluids in contact.
These results were obtained with colloidal suspensions brought in contact with their own solvent, for which interdiffusion can be neglected
on the time scale of the experiments.
We have shown that the presence of an effective interfacial tension rationalizes
the experimental findings satisfactorily: the injection rate dependence of the observed number of fingers is the one predicted for a finite
(positive) interfacial tension ($n_A\propto \sqrt{\dot{V}}$). Therefore, our results raise the question of whether the magnitude of interfacial
stresses mimicking capillary forces determine the morphology and the dynamics of viscous fingering patterns and whether fractal-like patterns
and Paterson scaling for $\lambda_{c}$ are observed only for very high capillary numbers, i.e. when interfacial stresses can be neglected.
Such stresses, namely the so-called Korteweg stresses discussed in section \ref{Korteweg}, are crucial in determining the onset and the
late-stage morphology of instabilities patterns in miscible fluids.
%Take home message%
A unified and solid framework allowing one to predict the onset and the evolution of viscous fingering
in miscible fluids remains elusive unless transient capillary forces are invoked. The latter depend on the physico-chemical nature of the fluids
in contact \cite{TruzzolilloPRX2016} and are not negligible in the case of sharp compositional mismatches.

\subsection{Stratified fluids under steady forcing: Kelvin-Helmholtz and Rayleigh-Taylor Instabilities}\label{KHRT}
Another interesting and ubiquitous configuration of fluids leading to hydrodynamic instabilities is sketched in Figure \ref{summary}-B.
When two fluids flow or lie one on top of the
other, their mutual interface may be unstable. When the fluids are in co-flow the instability is said to be of the Kelvin-Helmoltz type, while, if
they stay at rest, a Rayleigh-Taylor instability may arise.
The more general (and simple) case is that of two inviscid fluids, i.e. where inertia due to viscosity and dissipation can be neglected. For
both the Kelvin-Helmholtz and the Rayleigh-Taylor instabilities two inviscid miscible fluids of density $\rho_1$ and $\rho_2$ are separated by
an interface, let's say at $z=\zeta$. Hereafter we will assume no surface tension at the boundary between the fluids, and
that diffusion is slow compared to the time scale characterizing the onset of the instability. The interface $\Upsilon$ is defined by
\begin{equation}\label{surface}
    \Upsilon\equiv z-\zeta(x,y,t)=0,
\end{equation}
which evolves according to
\begin{equation}\label{motion_surface}
    \frac{d\Upsilon}{dt}=v_z-\frac{\partial \zeta}{\partial t}-v_x\frac{\partial \zeta}{\partial x}-v_y\frac{\partial \zeta}{\partial y}=0
\end{equation}
The governing equations are:
\begin{eqnarray}\label{gov_eq1}
  \frac{\partial v_x}{\partial x}+\frac{\partial v_y}{\partial y}+\frac{\partial v_z}{\partial z}=0 \hspace{0.5cm}\text{(Continuity equation)}\\
   \rho\frac{d \mathbf{v}}{d t}=-\nabla \Phi  \hspace{0.5cm} \text{(Euler equation)}
\end{eqnarray}
where $\mathbf{v}=(v_x,v_y,v_z)$ is the velocity field and $\Phi=p+\rho g z$. We impose the following boundary values for $\mathbf{v}$:

    \begin{equation}\label{laminar0}\mathbf{v}=\hat{x}\left\{
                 \begin{array}{ll}
                  U_1 \quad z\rightarrow + \infty\\
                  U_2 \quad z\rightarrow - \infty\\
                \end{array}
              \right.\end{equation}
and assume the basic flow given by
   \begin{equation}\label{laminar1}\mathbf{v}=\mathbf{V}\equiv\hat{x}\left\{
                 \begin{array}{ll}
                  U_1 \quad z > 0\\
                  U_2 \quad z < 0.\\
                \end{array}
             \right.\end{equation}
Equations \ref{motion_surface} and \ref{gov_eq1} can be solved in reciprocal space defined by the wave vector $\mathbf{k}\equiv(k,l)$,
leading to a dispersion relation that reads \cite{joseph_short-wave_1990}:

\begin{multline}\label{growthrate}
    \sigma(k,\beta)=-ik\frac{\rho_1U_1+\rho_2U_2}{\rho_1+\rho_2}\pm\\ \sqrt{\frac{k^2\rho_1\rho_2(U_1-U_2)^2}{(\rho_1+\rho_2)^2}-\frac{\beta g(\rho_1-\rho_2) }{\rho_1+\rho_2}},
\end{multline}
where $\beta=\sqrt{k^2+l^2}$. The real part of
$\sigma(k,\beta)$ is the growth rate of the mode associated to the wave numbers $k,l$.
The two-fluid system is unstable if the square root is strictly positive, i.e. when
\begin{equation}\label{cond_instab}
    k^2\rho_1\rho_2(U_1-U_2)^2>\beta g (\rho_1^2-\rho_2^2)
\end{equation}
We now draw the main conclusions on the instability to short wavelengths by inspecting Eqs \ref{growthrate} and
\ref{cond_instab} at large $k$. If $U_1\neq U_2$ and $k$ is large we have an instable eigenmode with growth rate

\begin{equation}\label{growth_KH}
    Re(\sigma)=k\left|\frac{U_1-U_2}{\rho_1+\rho_2}\right|\sqrt{\rho_1\rho_2}.
\end{equation}
This is known as the Kelvin-Helmholtz instability. If $U_1=U_2$ and $\rho_2>\rho_1$ (heavy fluid above) the unstable eigenvalue reads
\begin{equation}\label{growth_RT}
    Re(\sigma)=\sqrt{\frac{\beta g(\rho_2-\rho_1)}{\rho_1+\rho_2}},
\end{equation}

which is known as the Rayleigh-Taylor instability. Both the Kelvin-Helmholtz and Rayleigh-Taylor instabilities are
catastrophic short-wave instabilities of the Hadamard-type, as it can be deduced from Eqs. \ref{growth_KH}
and \ref{growth_RT} when $k$ tends to infinity. Although such results are strictly valid for inviscid fluids,
Funada and Joseph \cite{funada_viscous_2001}, have showed,
that for a vanishing surface tension such catastrophic behavior persists for the Kelvin-Helmoltz instabilities occuring between viscous fluids
in rectangular channels in absence of surface tension. By contrast a (finite) wave vector having maximum growth rate
appears when surface tension is strictly positive. It's worth pointing out that the analysis by Funada and Joseph neglects the no-slip condition
at the interface between the two fluids, resulting in a shear stress singularity at the interface and making their results
acceptable only for low viscosity fluids, where shear stresses can be neglected.

\begin{figure}[htbp]
 \begin{center}
  \includegraphics[width=7cm]{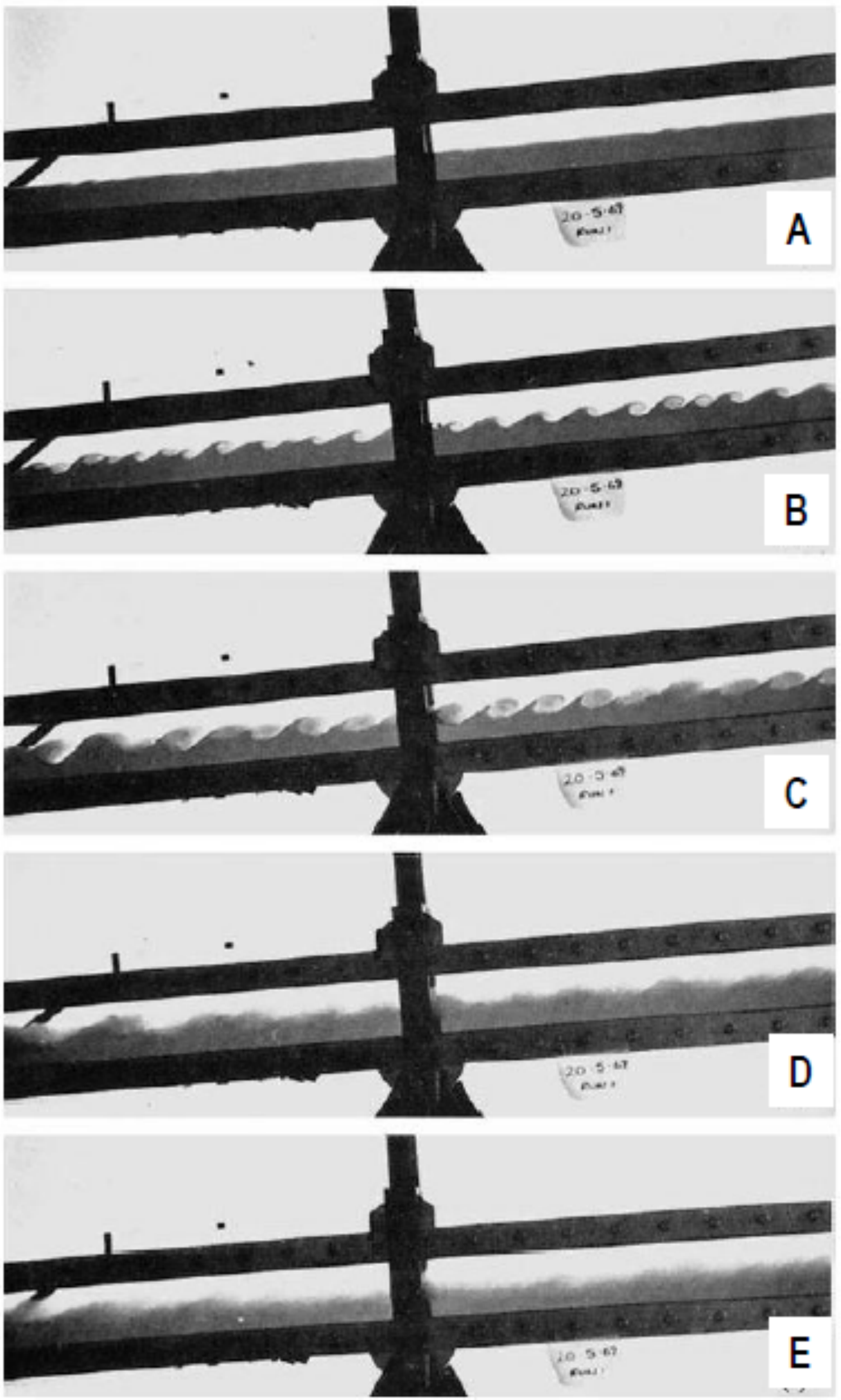}\\
  \caption{Laboratory demonstration of a Kelvin-Helmholtz instability. A long tank initially filled with two resting fluids
(pure and salt-saturated water) of different densities is tilted slightly. As the denser fluid sinks toward the bottom
and the lighter fluid rises toward the top, a counter flow is generated, and the interface develops unstable billows.
(adapted with permission from \cite{thorpe_experiments_1971})}\label{KH-lab}
\end{center}
\end{figure}

The results obtained with no surface tension are thus unphysical and in contrast with all experimental observations,
where one well-defined wavelength characterizes the onset of the K-H instability. The latter can be observed at very large scales,
e.g. for the wavy clouds seen in windy days (Figure \ref{summary}-B), or at the laboratory scale (Figure \ref{KH-lab}),
where the Kelvin-Helmoltz instability appears in channels filled with low viscosity miscible fluids flowing at different average speeds.

%Hence experiments and simple direct observation of natural phenomena suggests that Hadamard instabilities in idealized problems like the
%one we considered in this section, should not be considered as a result of no fundamental importance in the study of unstable multi-fluid
%systems. Their appearance shows that there is a kind of instability associated to short wavelengths which necessitate the action of certain
%stabilizing mechanisms which must be relevant for the hydrodynamics at low Reynolds numbers.

As for viscous fingering, there were attempts to regularize the Kelvin-Helhmoltz instabilities. A first strategy is to keep
the inviscid approximation, but to spread the vorticity over a finite layer. In this approximation, the
undisturbed velocity is initially continuous but the vorticity is discontinuous. Lord Rayleigh showed \cite{Rayleigh1880} that the finite vortex
layer is stable in both the long-wave and the short-wave limits and that the maximum growth rate occurs for wavelengths approximately 8
times the layer thickness. More recently Pozrikidis and Higdon \cite{pozrikidis_nonlinear_1985-1} argued that the amplitude of the disturbance
is relatively insensitive to the thickness of the finite vorticity layer and reaches a maximum value of approximately 20 $\%$ of the wavelength.
Although the non-zero width of the vorticity layer has been debated as a possible regularizing mechanism of K-H instabilities,
Joseph \cite{joseph_short-wave_1990} noted that the finiteness of the non-zero vorticity layer is not the key point of the regularizing mechanism.
Instead he proposed that the well-posedness of the problem is recovered just by lowering the order of the discontinuity: by considering two density-matched inviscid fluids separated by a flat interface and
velocity profiles continuous everywhere including at the interface, with constant but different vorticity above and below the interface, the shear flow becomes stable for any wave number, even when the surface tension is identically zero.
We emphasize that continuous velocity profiles are a most natural assumption for viscous fluids.
Indeed a sudden velocity jump at the interface
would imply an infinite (and unphysical) shear stress difference that, on the contrary,
must be zero under stationary conditions. For viscous fluids the stability problem stays always well-posed in the Hadamard sense:
shortwave lengths, being the most dissipative, are stabilized by viscosity
\cite{govindarajan_instabilities_2014-1} and the growth rate tends to zero for vanishing wavelengths.

\begin{figure}[htbp]
  \includegraphics[width=9cm]{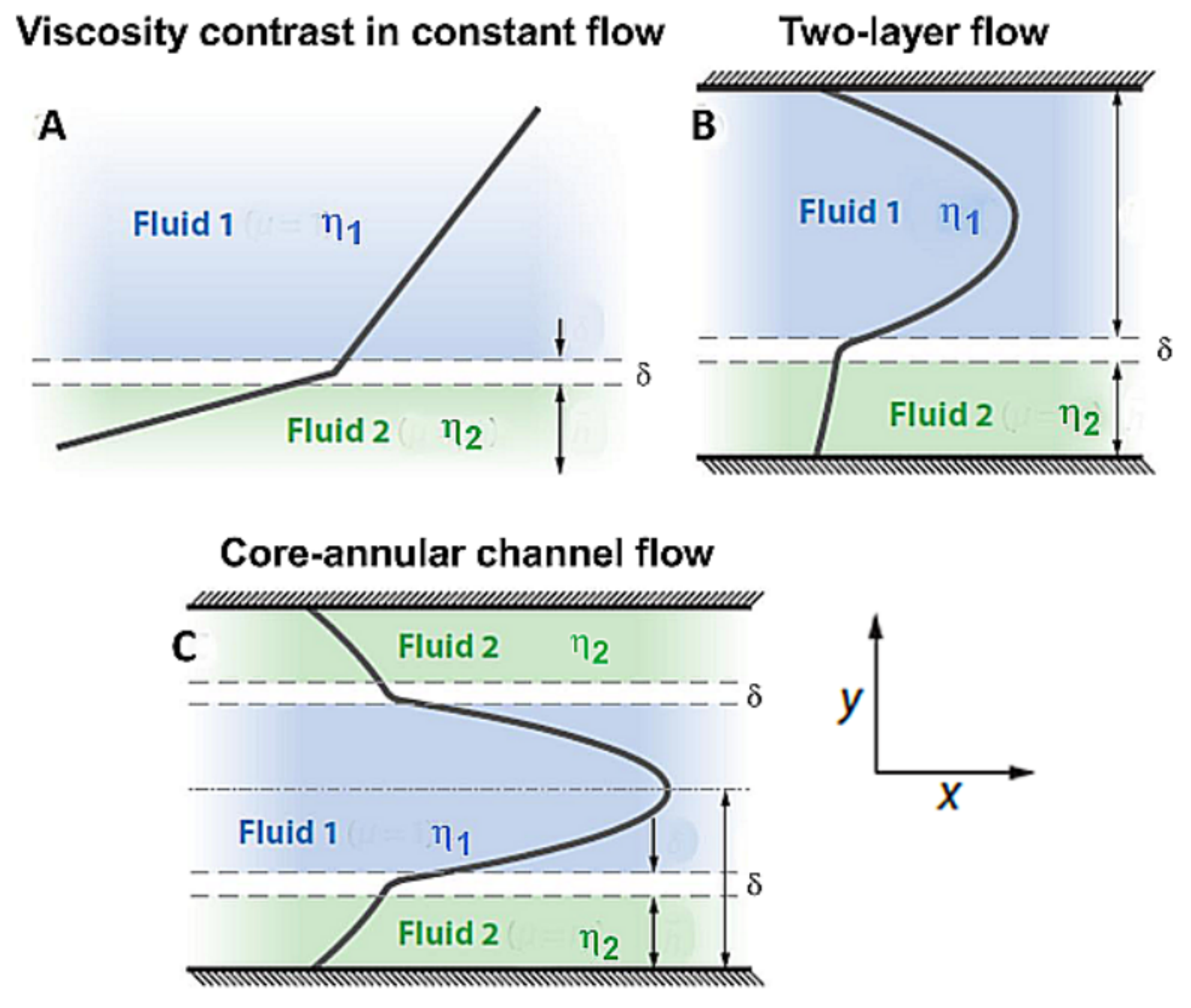}\\
  \caption{Velocity profiles in simple geometries, with viscosity variation $|\eta_1 - \eta_2|$ confined to a layer of thickness
  $\delta$. (A) An elbow profile created by a change in viscosity in constant shear-stress flow.
  (B) Two-layer flow. (C) Core-annular channel flow.}\label{stratified}
\end{figure}

An accurate analysis of the problem has been performed by Hooper and Boyd \cite{hooper_shear-flow_1983}
who considered the stability problem of the shear flow of superimposed immiscible viscous fluids with a linear velocity profile
(i.e. uniform vorticity $\nabla \times \mathbf{v}$) above and below a flat interface (Figure \ref{stratified}-A). They showed that this viscous
analogue of the inviscid Kelvin-Helmholtz problem is well posed even in the absence of surface tension, although
the flow is unstable at any Reynolds number.

The same kind of regularization via viscous dissipation can be found for the Rayleigh-Taylor instability \cite{rayleigh_theory_1896,joseph_short-wave_1990}.
Due to the general and ubiquitous nature of stratified flows, a lot of effort
\cite{Yih1967,yih_stratified_1969,Li1969,Hickoxc1971,hooper_shear-flow_1983,barmak_stability_2016}
has been devoted to the stability problem in more realistic situations, e.g. two fluids with different viscosities flowing between
two rigid plane-parallel boundaries separated by a finite distance. Three examples are given in Figure \ref{stratified}.

Figure \ref{stratified}-A shows the case analyzed by Joseph for inviscid fluids and by Hooper and Boyd for fluids with finite viscosities:
two co-flowing fluids with two uniform shear rates above and below the interface. Figure \ref{stratified}-B
shows the more classical case of a two-layer Poiseuille or Couette flow. The third example Figure \ref{stratified}-C sketches the case
of core-annular
flows where a less viscous fluid is flowing surrounded by the more viscous one.
In all the situations sketched in Figure 4, the interface is initially smooth when the fluids are set in rectilinear motion,
either by an applied pressure gradient (Poiseuille flow), or by the relative motion of the boundaries (Couette flow), a long-wavelength
instability arises, which persists at arbitrarily small values of the Reynolds number \cite{Yih1967,Li1969,Hickoxc1971}.
By considering perturbations containing all wavelengths, Barmak \emph{et al.} \cite{barmak_stability_2016} showed that without surface tension the
larger the wavenumber, the smaller the stable region in the parameter space of flow velocities.
They concluded that for sufficiently large wave numbers
no stable region exists, which is in agreement with the analytical asymptotic analysis for long wavelengths performed by
Yiantsios and Higgins \cite{yiantsios_linear_1988}.

\begin{figure}[htbp]
 \begin{center}
  \includegraphics[width=7cm]{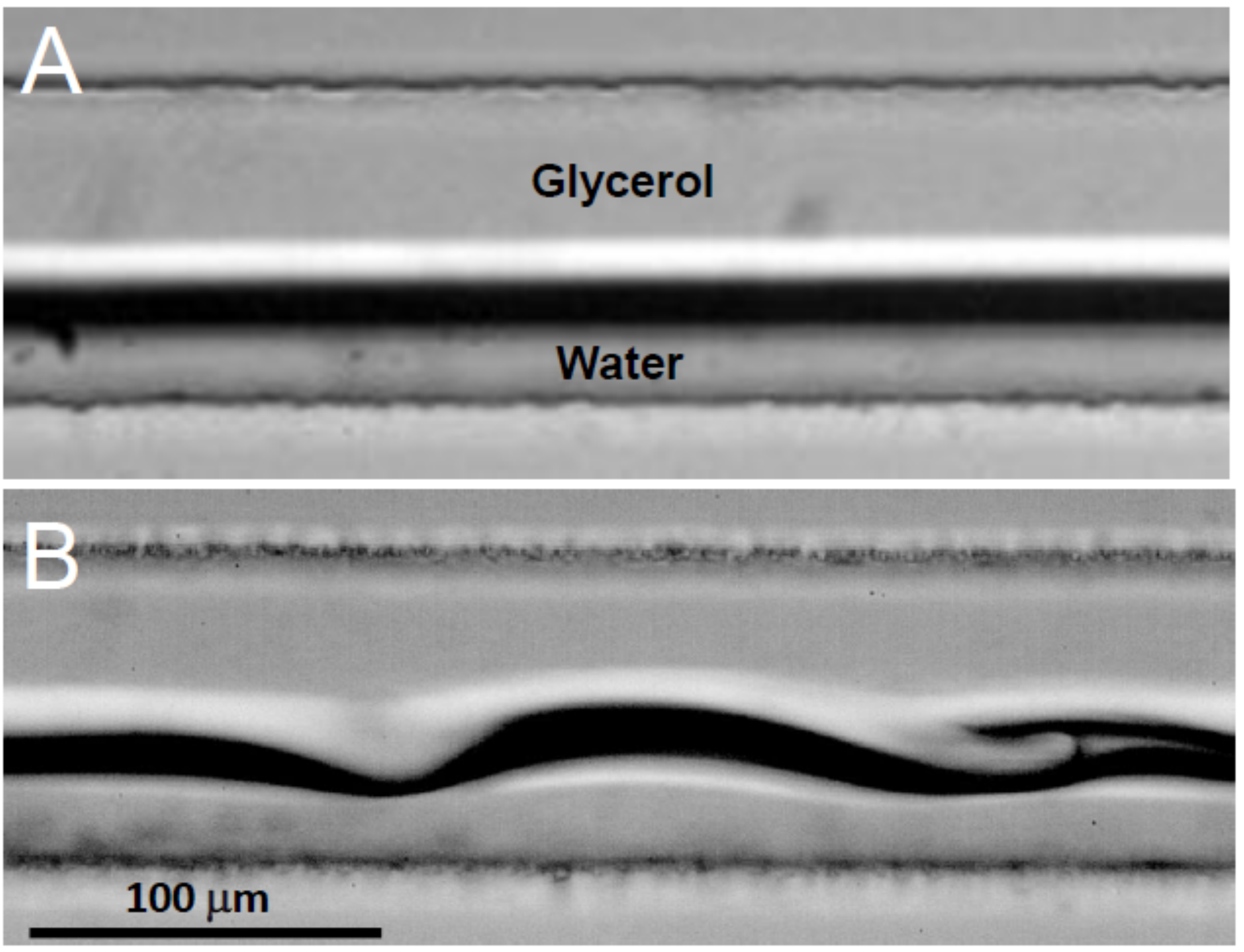}\\
  \caption{Co-flow of water and glycerol in a microchannel with squared section ($100 \mu m$ x $100 \mu m$).
  Panel A shows water and glycerol under a stable flow condition: $\dot{V}_{H2O}=40$ $\mu l$/min, $\dot{V}_{Gly}=0.625$ $\mu l$/min.
  Panel B: Shear flow instability observed when water and glycerol are injected at constant flow rates
  $\dot{V}_{H2O}=150$ $\mu l$/min, $\dot{V}_{Gly}=0.625$ $\mu l$/min.}\label{KH-water-gly}
\end{center}
\end{figure}

We conclude that when all wave numbers are considered, \emph{smooth stratified flow cannot be stabilized without surface tension}.
This is in stark contrast with experiments that we have recently performed \cite{Savorana2017}:
the stratified co-flow of two miscible fluids, namely glycerol and water
is stable for low enough Reynolds numbers, with a wave amplitude going to zero for a finite (positive)
flow rate of water, above which one single wavelength is selected (see Figure \ref{KH-water-gly}).
Whether this is due to stabilizing effects, e.g. energy dissipation, to effective capillary forces or to confinement, remains to be established.
For a detailed analysis of the stability of core-annular flows and other type of flows we refer the interested reader
to \cite{joseph_core-annular_1997,govindarajan_effect_2004,ranganathan_stabilization_2001,selvam_stability_2007}.

So far we have discussed the stability problem for miscible interfaces under a steady forcing.
However, hydrodynamic instabilities in miscible fluids arise also when the forcing is oscillatory, exhibiting distinct features with respect
to steady flow instabilities.
We will explore oscillatory forcing in the following two sections.

\subsection{Oscillatory forcing I: Faraday instability}\label{Faraday-sec}
The onset of flow patterns in density-stratified fluid layers, subject to an oscillatory forcing perpendicular to the fluid interface, is
known as the Faraday instability \cite{faraday_peculiar_1831}. While Faraday instabilities can generally occur in both miscible and immiscible
fluid layers, the phenomena owe their origins to different transport mechanisms as we discuss shortly. In the case of immiscible liquids, the
destabilization of the static layers is manifested by the transverse variation in interfacial elevation and occurs due to the resonance between the
imposed frequency and the natural frequency of the system. The latter depends on the density and viscosity difference between the fluid layers and
the surface tension at their boundaries. The surface tension and viscous dissipation effects stabilize all the perturbations below
a critical limit of the oscillation parameters (frequency and amplitude). Beyond this limit, interfacial deformation takes the shape of a standing
wave with a defined wavelength and, as we will shortly see, instability is not affected by ill-posedness in the Hadamard sense
\cite{landau1989mécanique,zoueshtiagh_experimental_2009,diwakar_faraday_2015}.

Whether or not miscible fluids are affected by capillary forces, it's worth to approach the problem with the very general case of capillary-gravity waves.
Let us consider (Figure \ref{summary}-C) a container partly filled with a Newtonian fluid with density $\rho$, moving up and down in a purely
sinusoidal motion of angular frequency $\Omega$ and amplitude $A$, so that the forcing acceleration is $\Omega^2A \cos(\Omega t)$. In the reference
frame moving with the vessel, the fluid (vertical) acceleration is $G(t) =g-\Omega^2A \cos(\Omega t)$, with $g$
the acceleration due to gravity in the laboratory (Galilean) frame and $t$ the time.
Let $\mathbf{x} = (x_1, x_2)$ and $y$ be the horizontal and upward vertical Cartesian coordinates comoving with the vessel, respectively.
Ordinates $y = -d$, $y = 0$ and $y = \zeta(\mathbf{x}, t)$ correspond to the horizontal impermeable bottom, the liquid level at rest
and the impermeable free surface, respectively. The Fourier transform of the latter is $\bar{\zeta}(k, t) = \int\int\zeta(\mathbf{x},t)e^{-i\mathbf{k}\cdot\mathbf{x}}d\mathbf{x}$,
where $\mathbf{k}$ is the wave vector with $k = |\mathbf{k}|$.
For surface waves of vanishingly small amplitude, $\bar{\zeta}$ is described by a damped Mathieu equation
\cite{benjamin_stability_1954,ciliberto_chaotic_1985}
\begin{equation}\label{Mathieu}
    \bar{\zeta_{tt}}+2\psi\bar{\zeta_{t}}+\omega_0^2(1-F\cos(\Omega t))\bar{\zeta}=0
\end{equation}
where $\psi = \psi(k)$ is the viscous attenuation that originates from the bulk viscous dissipation and the
viscous friction with the bottom in the case of a shallow fluid, $\omega_0 = \omega_0(k)$ is the angular frequency of linear
waves without damping and without forcing, and $F = F(k)$ is a dimensionless forcing.
For a non-vanishing surface tension $\Gamma$ and finite depth $d$, the solution to equation \ref{Mathieu} are capillary-gravity waves
characterized by the dispersion relation \cite{landau1989mécanique}:
\begin{equation}\label{dispFaraday}
    \omega_0^2(k)=(gk+\frac{\Gamma k^3}{\rho})\tanh(kd)
\end{equation}
The damped Mathieu (eq.\ref{Mathieu}) only holds for infinitesimal waves and neglects nonlinear effects that play a crucial role for steep waves.
Nonetheless it allows one to study some fundamental physical properties of the Faraday waves \cite{rajchenbach_faraday_2015}.
Rajchenbach \emph{et al.} have shown that the wavenumbers selected at the instability
onset depend uniquely on the forcing frequency $\Omega$, the viscous attenuation $\psi$, the depth of the liquid $d$ and its viscosity $\eta$.
In the two opposite limits of infinite depth and shallow liquid they obtained
\begin{equation}\label{wave1}
    2\omega_0=\Omega/2+\sqrt{\Omega^2/4-16\psi^2} \hspace{1cm} d\rightarrow \infty
\end{equation}
\begin{equation}\label{wave2}
    \omega_0=\Omega/2+16\eta/\rho d^2 \hspace{1cm}  \mbox{shallow liquid}.
\end{equation}

Equations \ref{dispFaraday}-\ref{wave2} show that the idealized problem in absence of surface tension is well posed:
for any frequency of the forcing it is possible to find a Faraday wave with strictly positive wavelength. Howaver a general linear stability
analysis of miscible systems is complicated by one significant challenging issue: the problem is time dependent, since diffusion takes place
in the forcing direction and the corresponding linear system ceases to be described by a Mathieu-type equation. This issue has been tackled only
recently \cite{zoueshtiagh_experimental_2009,diwakar_faraday_2015} by considering a frozen-time approximation wherein the
concentration profile is assumed to remain frozen during the evolution of the instability. This makes the resulting linear stability analysis
feasible, as shown by Narayanan and coworkers \cite{zoueshtiagh_experimental_2009,diwakar_faraday_2015}, who studied numerically and experimentally
the onset of the Faraday instability in miscible fluids. Within this approximation, the evolving concentration is considered to be frozen
during the growth and decay of the perturbations. This eliminates all the non-harmonic components of the evolution from the linear analysis and
is justifiable when the time scale of diffusion is much greater than the period of the parametric forcing, $\tau$.
As experimental systems the authors considered two different pairs of fluids: de-ionized water and salt-saturated water, and
pairs of silicone oils with different densities. In all the experiments, the onset of the Faraday instability was
studied by fixing the oscillation amplitude $A$ and by gradually increasing the frequency $f$, until the formation of a wave at the interface
was observed.

The motion of the observed standing waves was characterized by an oscillation frequency $\omega_0 =\Omega/2$.
This result, obtained both experimentally and numerically, is a further evidence
that the problem can be viewed as a classical Faraday problem, since it is consistent with Eq. \ref{wave1} in the
limit of negligible viscous dissipation.
Furthermore, in analogy with immiscible fluids, the wavelength increases upon increasing the mean viscosity of the two-fluid system or, more
precisely, when increasing the Schmidt number $Sc$, the latter being the ratio between the kinematic viscosity of the heavy fluid and the mass
interdiffusion coefficient.
\begin{figure}[htbp]
  \includegraphics[width=8.5cm]{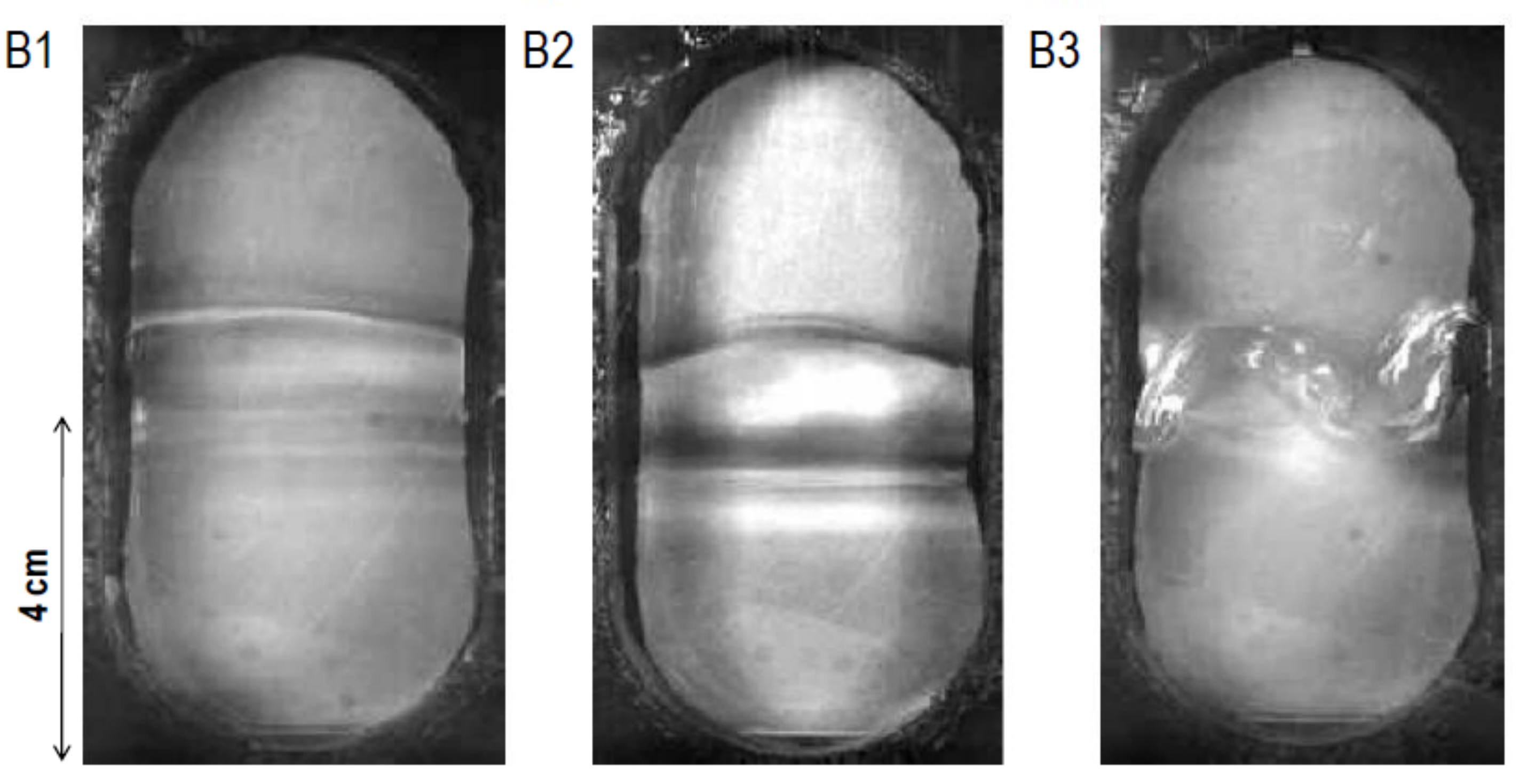}\\
  \caption{Influence of the waiting time $t$ on the Faraday wavelength.
  Side view of pure water-brine system under vertical vibration with amplitude $A=10$ cm,
  frequency $\Omega$ =9.42 rad/s and waiting times $t$=5 min (B1), $t$=20 min (B2) and $t$=45 min (B3)(adapted with permission from \cite{zoueshtiagh_experimental_2009})}
  \label{Faraday}
\end{figure}

In spite of these analogies, Faraday instabilities in miscible fluids are unique in that
the wavelength of the instability patterns at the instability onset may decrease (as shown in figure \ref{Faraday}) or increase as the time waited before imposing the external oscillations increases.
Such phenomenon is controlled by the thermo-physical properties of the fluid layers.
In particular Diwakar \emph{et al.} \cite{diwakar_faraday_2015} showed that the parameter dictating the wavelength selection and its dependence
on the waiting time is the Rayleigh number, defined as $Ra_T=\beta_c g \Delta C (DT)^{\frac{3}{2}}/D \nu$,which measures the buoyancy force relative to the viscous dissipation.
Here $\beta_c$ is the solute thermal expansion coefficient, $g$ the gravity acceleration, $\Delta C=C_1-C_2$ the initial concentration difference of
fluid 1 and 2 within the bulk volumes, $T$ the temperature, $D$ is the mass interdiffusion coefficient and $\nu$ the kinematic viscosity of the
heavier fluid.
These authors argued that the most unstable wavelength increases (decreases) with waiting time for small (large) Rayleigh number. Up to our
knowledge, only the large Rayleigh number behavior has been tested \cite{zoueshtiagh_experimental_2009} while the prediction for low Rayleigh
numbers is still unverified.

\subsection{Oscillatory forcing II: Standing Kelvin-Helmholtz waves}\label{KHOsc}
Faraday waves are not the only instabilities observed when two fluids are subject to vibrations.
Vibrations normal to the interface differ qualitatively from horizontal vibrations, tangential to the interface; the former affect the fluid
by changing the effective gravity and, thus, the hydrostatic pressure gradient.
The latter, by contrast induce a sheared flow altering the interface
stability, like in the Kelvin-Helmholtz case. As for the Saffman-Taylor and the Faraday instability we first discuss the immiscible
in the zero surface tension limit. For two superimposed immiscible liquids the effect of horizontal vibrations has been examined both
experimentally \cite{wolf_dynamic_1970,ivanova_interface_2001,jalikop_steep_2009,yoshikawa_oscillatory_2011,yoshikawa_oscillatory_2011-1} and
theoretically \cite{lyubimov_development_1987,khenner_stability_1999}, showing that an initially flat interface may become unstable and that
non-propagating periodic waves may develop at the interface. Following Wunenburger \emph{et al.} (1999), these waves, which are stationary in the reference
frame of the vibrated cell, are referred to as "frozen waves".
By means of linear stability analysis Lyubimov and Cherepanov \cite{lyubimov_development_1987} gave the criteria for the interface stability for
interfaces with positive interfacial tension. They predicted that the interface between two liquid layers of equal thickness $H$,
densities $\rho_1$ and $\rho_2$ and interfacial tension $\Gamma$, becomes unstable to a sinusoidal disturbance in the limit of high
frequencies $\Omega\gg H^2/\nu$ (i.e. where the period of vibrations is much smaller than the viscous time scale), for vanishing amplitudes
of the forcing $A\ll H$, and when the velocity amplitude exceeds a critical value such that the velocity difference $\Delta U$
between the two fluids fulfills the inequality
\begin{equation}\label{h-threshold}
\Delta U^2>\frac{2(\rho_1+\rho_2)}{\rho_1\rho_2}\left(\frac{2\pi\Gamma}{\lambda}+\frac{(\rho_1-\rho_2)g\lambda}{2\pi}\right)
\tanh\left(\frac{H2\pi}{\lambda}\right)
\end{equation}
with a critical wavelength \cite{wunenburger_frozen_1999}
\begin{equation}\label{lambda-threshold}
\lambda_{cr}=2\pi\left[\frac{\Gamma}{\rho_1-\rho_2}\right]
\end{equation}
appearing for the lowest value of $\Delta U$ fulfilling the inequality \ref{h-threshold}.
\begin{figure}[htbp]
  % Requires\usepackage{graphicx}
  \centering{\includegraphics[width=8cm]{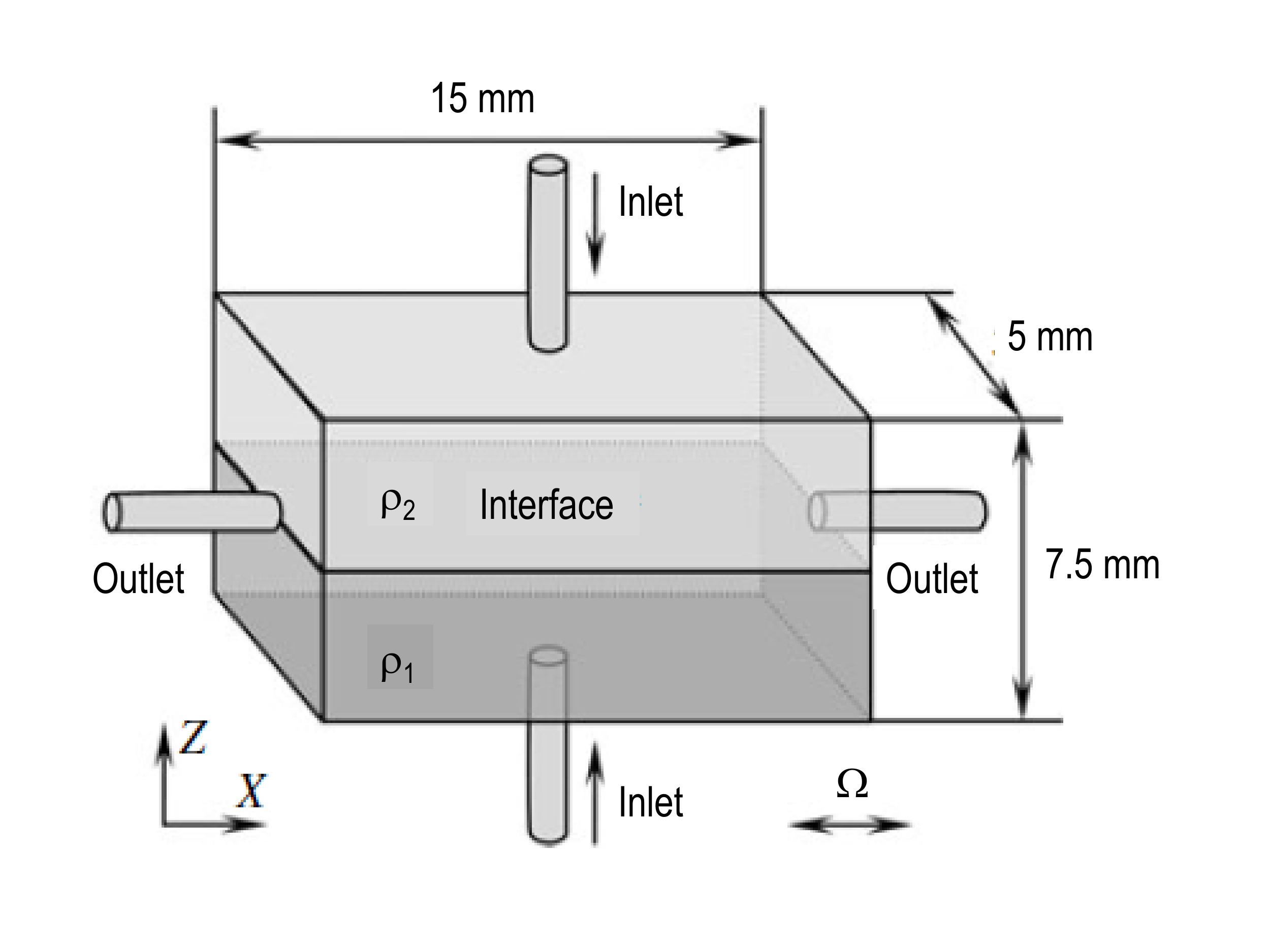}}\\
 \centering{\includegraphics[width=8cm]{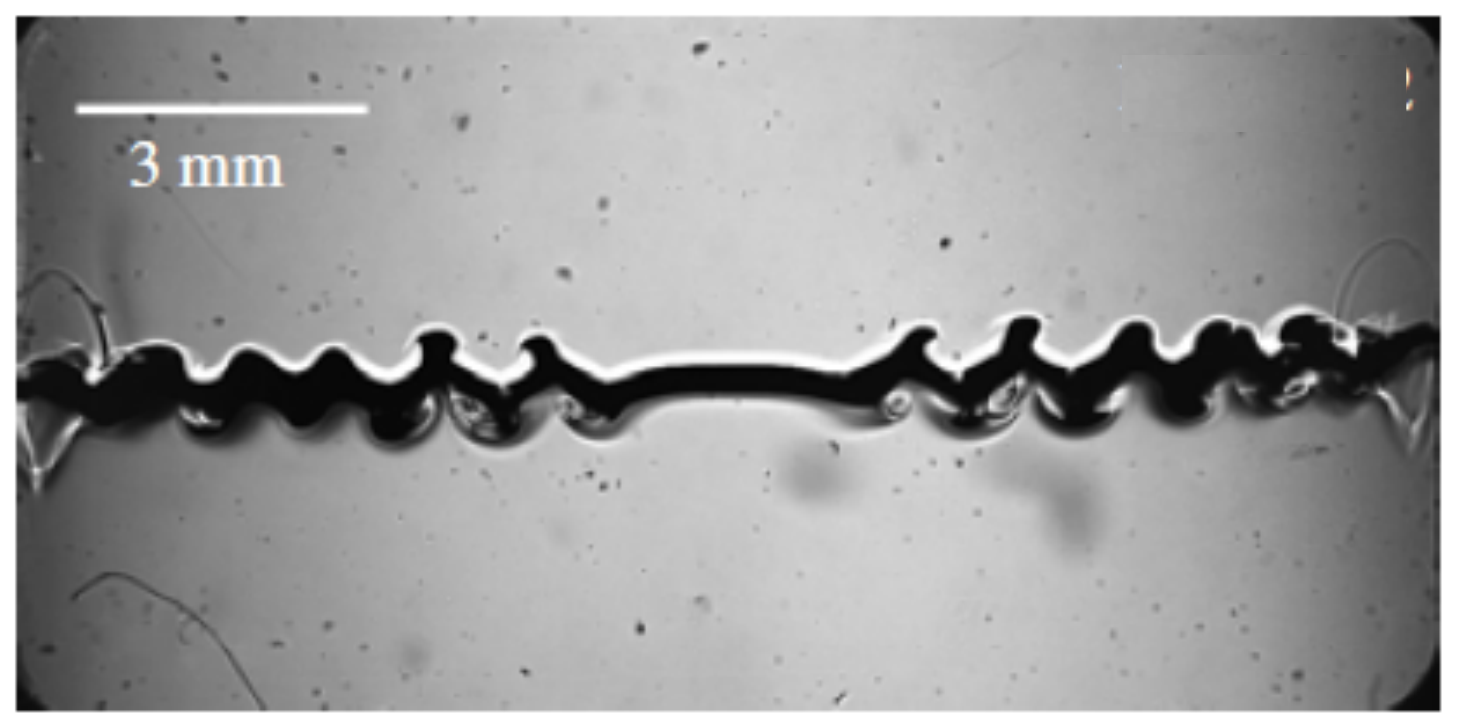}}\\
  \caption{Upper Image: Experimental cell used in ref. \cite{gaponenko_dynamics_2015}. Both liquids are injected into the cell simultaneously
  by two identical syringe pumps through the orifices in the bottom (denser liquid) and top (lighter liquid) walls. The excess of liquid
  leaves the cell through the orifices located in the end walls (adapted with permission from \cite{gaponenko_dynamics_2015}).
Bottom Image: Interfacial pattern visualized when two super-critical (miscible) water-isopropanol mixtures are brought in contact under horizontal
vibration. Experiments are performed $1.7$ s after loading the cell, for horizontal oscillations at
frequency $f=$22.5 Hz and amplitude $A=$3.6 mm. Adapted with permission from \cite{gaponenko_dynamics_2015}.}\label{Gaponenko}
\end{figure}
Interestingly, Eq. \ref{h-threshold} is analogous to the dispersion relation of the basic instability mode of Kelvin-Helmholtz (K-H)
type for two counter flows of infinitely thick layers ($H\rightarrow\infty$) when $(U_1-U_2)/\sqrt{2}$ is regarded as the velocity
difference between the two fluids in the ideal K-H case (Equation \ref{cond_instab}).
Horizontal vibrations applied to a cell with two superimposed liquids generate a horizontal pressure gradient that induces oscillatory
shear flows. The peculiarity is that, as a result of the harmonic change in the flow direction, the wave remains on average in the same place,
as its profile is frozen in the reference frame of a vibrating container. As for the classic Kelvin-Helmholtz instability in the absence
of surface tension, the critical wavelength that first appears is zero (see Eq.\ref{lambda-threshold} with $\Gamma$=0): the problem is Hadamard
ill-posed and does not conform to experiments, as we see hereafter.

Experiments with miscible fluids are associated with severe experimental difficulties in forming a sharp interface, especially for fluids with
similar viscosities, and the system of two superimposed liquids has to be newly created after each experimental test.
The only known vibrational experiments, up to our knowledge, were conducted by Legendre \emph{et al.} \cite{legendre_instabilites_2003}
with water and glycerol and by Gaponenko \emph{et al.} \cite{gaponenko_dynamics_2015} with mixtures of water and isopropanol.
In both cases, finite unstable standing waves were observed at a critical velocity amplitude.
One of the features that highlights the difference between immiscible and miscible fluids is the shape of frozen waves.
Gaponenko \emph{et al.} have argued that (almost) vanishing interfacial tensions and negligible viscous effects are responsible for
triangular-shaped waves (Figure \ref{Gaponenko}), while for immiscible fluids frozen waves are sinusoidal.
In principle, the standing waves of the oscillatory K-H instability are not affected by the viscous component of the stress tensor.
However Legendre \emph{et al.} \cite{legendre_instabilites_2003} (when using glycerol ($\eta_1=1000$ mPa s) and water ($\eta_2=1$ mPa s))
have showed that sine waves do appear. Gaponenko \emph{et al.}
explained this surprising result by arguing that 'frozen' waves are in fact slightly
moving, such that the viscous part of the stress cannot be discarded completely:
viscous stresses tends to stabilize short wavelengths, smoothing out the disturbances and
determining the sine waves observed by Legendre \emph{et. al}.

Regardless the wave shape, this interfacial instability in miscible liquids sets in beyond a well-defined threshold.
Upon increasing the value of the oscillatory
speed above the critical value, the amplitude of the perturbation continuously grows, forming a saw-tooth frozen structure similar
to that shown in Fig. \ref{Gaponenko} for the water-isopropanol mixture.
By means of optical observations, Gaponenko \emph{et al.} have built a vibrational
forcing map delineating the regions of stability and instability.
The comparison of these results with the theoretical predictions by Lyubimov and Cherepanov (1987)
\cite{lyubimov_development_1987} for immiscible liquids has shown that the inviscid model (Eq. \ref{h-threshold})
significantly underestimates the threshold of the instability, while experiments and inviscid theory are in reasonable
agreement for high frequencies $\rho_1\Omega H^2/\eta_1>350$. Finally, it is interesting to note that, in an attempt to rationalize their finding,
Gaponenko \emph{et al.} made use of Cahn-Hilliard-Navier-Stokes equations \cite{kheniene_linear_2013} containing a positive capillary coefficient,
thus assuming a small non-zero interfacial tension between the fluids. This (debated)
assumption appears once again in the context of hydrodynamic instabilities, demonstrating its control role in the
stability analysis of miscible interfaces.

\section{Korteweg stresses: an interfacial tension for miscible fluids}\label{Korteweg}

\subsection{Experimental evidences}
We have seen that the assumption of zero capillary forces at miscible boundaries rises a series of problems
when analyzing hydrodynamic instabilities and when attempting to interpret experimental results. Thus, the question naturally arises of
\emph{what kind of forces actually characterize miscible interfaces and drive their stability.} \emph{Do miscible liquids behave as they
were immiscible, at least before they intermix?}\emph{Does an effective interfacial tension stabilize hydrodynamic instabilities?}

Determining the conditions under which boundaries between miscible fluids are stable would require the full knowledge of the stress
tensor in the region where sharp concentration, density, viscosity or temperature gradients characterize the system.
This is clearly a very challenging task. While for immiscible fluids the absence of mass transfer and the existence
of a stabilizing surface tension help in determining such conditions,
for miscible fluids the stability problem is still a debated one. A certain number of experiments \cite{Truzzolillo2017}
unambiguously suggests that a
mechanism beyond viscous dissipation regularizes (in the Hadamard sense) the stability problem.
Dynamical effects that can arise in thin mixing layers where the compositional gradients are large can be considered as a good candidate to
explain the behavior of miscible interfaces.
Such possibility was recognized in discussions given by Korteweg as early as 1901 \cite{Korteweg1901}, where he proposed a constitutive equation
that included stresses induced by gradients of composition that mimic surface tension.
A short yet fascinating historical survey of the elusive interfacial tension in miscible liquids is given by Joseph \cite{Joseph1990},
in the context of his own work on the presence of sharp interfaces during the slow motion of rising drops of water in a background miscible fluid.
He reports many
experimental evidences supporting the existence of interfacial stresses in miscible fluids, dating back to the first
considerations by M.J. Bosscha, the first author to invoke the existence of appreciable capillary forces in the layer between two liquids miscible in all proportions.
As reported by Korteweg \cite{Korteweg1901}, Bosscha discussed some observations of the slow motion of acid or salt aqueous solutions in water or
in less concentrated solutions, and concluded that some kind of capillary forces had to be at work in order to explain
the formation of connected rings.
A similar description of drops of miscible and immiscible liquids occurring during the evolution of falling
drops into vortex rings has been reported much more recently \cite{baumann_vortex_1992}.
In a similar context, a membrane which continuously connect the torus that forms during the evolution of a falling drop of 9/10
glycerin/water mixture in a 3/2 glycerin/water mixture has been observed by Arecchi \emph{et al.} \cite{0295-5075-9-4-006}.
Such a membrane is hard to be explained without acknowledging some kind of interfacial tension.
In addition to these observations, there were many attempts to perform a more quantitative analyses, by measuring Korteweg stresses
at miscible interfaces and by determining their role (stabilizing or destabilizing).

In the following, we will mention only few significant experiments shedding light on the existence of
Korteweg stresses and employing three different techniques or strategies.
For a more detailed discussion, we refer the reader to our recent review \cite{Truzzolillo2017}.
Among all standard tensiometric methods, spinning drop tensiometry (SDT) has been quite systematically used
\cite{Petitjeans1996,Pojman2006,Pojman2007} to probe miscible interfaces,
despite the large equilibration time required to perform a measurement ($\sim 10^2$ s).
This precludes the possibility of probing stresses in the absence of diffusion. In spite of this limitation, all
spinning drop experiments performed on pairs of miscible liquids suggested that stresses mimicking a positive effective interfacial tension (EIT)
do exist.

P. Petitjeans~\cite{Petitjeans1996} was the first to perform such kind of investigation. He measured the EIT between water and aqueous
solutions of glycerine. After approximately 100 seconds drops of water reach a quasi-steady diameter when rotated in a capillary filled with
pure glycerin. By analyzing the drop shape, he obtained a positive EIT of 0.58 mN/m. He then replaced pure glycerine by water-glycerine mixtures
showing that the EIT is an increasing function of the mass fraction $\varphi_{gly}$ of glycerine.
Another series of very accurate SDT experiments have been performed by Pojman \emph{et al.} \cite{Pojman2006} on a near-critical water/isobytiric
acid (IBA) mixture. To demonstrate that an effective interfacial tension exists, Pojman and coworkers followed the evolution of a IBA-rich
drop embedded in a IBA-poor phase, above the upper critical solution temperature (UCST) of the mixture, i.e. in the one-phase region where
the two liquids intermix. When decreasing the angular speed $\omega$ of the apparatus, the drop changed its shape and reduced its interface area,
a clear evidence for a positive interfacial tension. In an analogous work, the same group \cite{Pojman2007} measured positive interfacial
stresses in monomer/polymer mixtures, where monomers and polymers shared the same chemistry. This suggests that Korteweg stresses
can arise even in the sole presence of gradients of configurational entropy, i.e when the two fluids in contact are
distinguished by a different connectivity of their constituents.

The second method to probe the EIT is light Scattering (LS) which has been used to follow the time evolution of the
effective tension in near-critical mixtures.
One of the most inspiring experiments was conducted by May and Maher~\cite{May1991}. They considered a system of mutually saturated layers
of isobutyric acid (IBA) and water. The two fluids have an UCST $T_c = 27.6~^\circ$C, above which
they are fully miscible, while they are immiscible below $T_c$. The authors performed dynamic light scattering experiments to obtain
the interfacial tension below $T_c$, by measuring the relaxation rate of the capillary waves at the interface between the two (immiscible)
fluids. They then ramped  up quickly the temperature just above $T_c$ and observed that the solution still displayed a behavior indicative
of the existence of an actual interface, with an EIT that relaxed with time as the fluids mixed together.

Cicuta \emph{et al.}~\cite{cicuta01} used static LS to investigate the non-equilibrium fluctuations occurring at the interface between two miscible
phases of a near-critical aniline-cyclohexane mixture during a free-diffusion process. They model the region separating the two fluids as two
thick ``bulk'' layers, within which a low and linear gradient of concentration exists~\cite{Cicuta2000}, located on both sides of a thin
interface layer characterized by a steep concentration gradient. The complex interplay between bulk and interface layers demonstrated
by the work of Cicuta \emph{et al.} highlights the difficulty of disentangling the two contributions, which ultimately limits the usefulness of
LS as a probe of the EIT.

The third and more recent strategy adopted to measure interfacial stresses at miscible interfaces takes advantage of the visualization of
instabilies and fluid patterns arising when two miscible fluids are under flow.
Schaflinger \emph{et al.} \cite{schaflinger_interfacial_1999} observed the time evolution of a falling drop of a suspension of glass beads in a
reservoir of their own solvent (glycol). In the case of relatively large droplets with a diameter $D>4$ mm, an emanating tail existed over
the whole height of the container. The column of liquid bulged with a wavelength that compares very well with theoretical prediction by
Tomotika \cite{tomotika_instability_1935}, who studied the instability of a cylindrical thread of a viscous liquid surrounded by another
viscous fluid. Capillary forces usually cause and tune such instabilities. For this reason, the authors concluded that the falling suspension
droplet exhibited an apparent interfacial tension ranging from about $2\cdot 10^{-2}$ mN/m to $1.6\cdot 10^{-1}$ mN/m.
They also concluded that an interfacial tension exists between the suspension and the solvent from the bulging of long
cylindrical threads.

As shortly discussed in section \ref{KHOsc}, Gaponenko \emph{et al.} \cite{gaponenko_dynamics_2015} recently presented the experimental evidence of
the existence of an interfacial instability associated to a Kelvin-Helmoltz instability between two miscible liquids of similar (but non-identical)
viscosities and densities under horizontal vibration. By measuring the critical wavelength at the onset of the instability,
they were able to measure the interfacial tension between mixtures of water and isopropanol of different
compositions.

The visualization of radial Saffman-Taylor instabilities falls into the same category of experiments. It has have been used
very recently to measure interfacial stresses between colloidal suspensions and their own solvent
\cite{Truzzolillo2014,Truzzolillo2015,TruzzolilloPRX2016}.
We showed that the effective tension between the suspensions and their solvent is dramatically affected by the microstructural details of the
fluids: a quadratic dependence on the
concentration characterizes (linear or crosslinked) polymer suspensions, while a much sharper dependence on volume fraction has been observed
for compact particles \cite{TruzzolilloPRX2016}. In the same work, we have proposed a new phase field formulation that is based on a frozen time
approximation, valid when diffusion is slow with respect to the time of the experiments. Among our recent results, we believe that it is instructive
to recall the analytical form obtained for the effective interfacial tension arising in the simple case of a symmetric binary mixture
in presence of a compositional gradient. This is the simplest case of a boundary between two miscible molecular fluids, yet it gives the
opportunity to discuss the main features of interfacial stresses at miscible boundaries. We shall do it in section \ref{lattice}.
Before doing so, we review the two approaches originally used to introduce Korteweg stresses: the fluid-dynamic formulation (sec. \ref{fluidK}) and
the thermodynamic approach (sec. \ref{SG-binary})

\subsection{Fluid-dynamic formulation}\label{fluidK}
The original formulation of capillary forces in miscible fluids dates back to the work of D. Korteweg \cite{Korteweg1901}
and was inspired by the theory of capillarity by van der Waals \cite{vanderwaals1894}. The latter argued
that the discontinuity at the interface between a liquid and its vapor is only apparent and that there is a layer of transition of thickness
much larger than the molecular interaction range.
For this problem, Korteweg proposed a continuum approach for a model compressible fluid driven by a "standard" stress $T^N$ of
the usual Navier-Stokes type, plus a "non-standard" part $T^K$ depending only on the density derivatives:

\begin{equation}\label{Navier-Stokes1}
    \rho\frac{Du}{Dt}=\nabla\cdot (T^N+T^K)
\end{equation}
where $\rho$ and $u$ are respectively the density and the velocity of the fluid,

\begin{equation}\label{Stress_Navier}
 T^N_{ij}=-p\delta_{ij}+\eta_1\left(\frac{\partial u_i}{\partial x_j}+\frac{\partial u_j}{\partial x_i}-\frac{2}{3}\delta_{ij}\frac{\partial u_k}{\partial x_k}\right)+\eta_2\delta_{ij}\frac{\partial u_k}{\partial x_k}
\end{equation}
and

\begin{equation}\label{Stress_Korteweg}
 T^K_{ij}=\left(\hat{\alpha}\nabla^2\rho+\hat{\beta}\nabla\rho\cdot\nabla\rho\right)\delta_{ij}+\hat{\delta}\frac{\partial \rho}{\partial x_i}\frac{\partial \rho}{\partial x_j}+\hat{\gamma}\frac{\partial^2 \rho}{\partial x_i\partial x_j}.
\end{equation}
Here the coefficients $\eta_1$ and $\eta_2$ are the shear and bulk viscosities, respectively \cite{landau1989mécanique}, and
$\hat{\alpha}$, $\hat{\beta}$, $\hat{\gamma}$ and $\hat{\delta}$ are functions of the local density $\rho$ and the temperature $T$. All the
coefficients appearing in $T^K$, in the absence of an appropriate molecular theory, must be determined experimentally, including their algebraic
sign. Korteweg considered the layer separating two miscible liquids of different density and calculated the difference of
the normal stresses acting on the boundary. He found quite strikingly that this difference is proportional to the mean curvature $\kappa$,
similarly to the immiscible case in presence of capillary forces. More precisely he showed that, for a spherical shell separating two fluids
with different densities and a mixing boundary characterized by a radial gradient $\frac{\partial\rho}{\partial r}$,
the pressure jump across the shell can be written as:

\begin{equation}\label{Korteweg-capillary}
\Delta \Pi=-\frac{2}{R}\int_0^{\infty}\left(\hat{\delta}+\frac{\partial\hat{\gamma}}{\partial\rho}\right)\left(\frac{\partial\rho}{\partial r}\right)^2dr=-\frac{2\Gamma_e}{R}
\end{equation}
where $r$ is the radial coordinate, $R$ the inner radius of curvature of the shell and $\Gamma_e$ an effective interfacial tension.
Equation \ref{Korteweg-capillary} is rigourously valid in the case of compressible fluids and is analogous of the Young-Laplace equation derived
by the classical theory of capillarity \cite{Rowlinson82}, with the surface tension coefficient having the same square gradient form derived by
Van der Waals \cite{vanderwaals1894} for immiscible phase-separated fluids at equilibrium.

Many years later, Joseph \cite{Joseph1990} has reconsidered the equations of fluid dynamics of two incompressible miscible liquids with
gradient stresses. He further assumed that the density of such incompressible fluids depend on concentration and temperature and that the
velocity field $u$ is in general not solenoidal, $\nabla \cdot u \neq 0$. Joseph pointed out that Korteweg stresses do not enter into the
stress jump across a flat mixing layer while, akin to interfacial tension, they do contribute to the pressure jump across a miscible interface
during the spreading of a spherical diffusion front. It's worth reporting here Joseph's expression for glycerine and water at
$T=20$ $^{\circ}$C \cite{joseph_tension-joseph.pdf_1996}
\begin{equation}\label{Pressure-Jump}
\Delta \Pi =\frac{2}{r_0}\sqrt{\frac{D}{t}}\left[-164.5\frac{\hat{\delta}}{D}-428.7\right],
\end{equation}
where $D$ is the diffusion coefficient of water in glycerine and $r_0$ is the initial radius of a spherical insertion of glycerin
in an infinite reservoir of water.
Such expression, once again, is reminiscent of the Laplace equation with a time dependent tension $T(t)$,
i.e $\Delta \Pi=\frac{2T(t)}{r_0}$. Note that $T(t)$ can be either positive or negative. Indeed there are two terms
in Eq. \ref{Pressure-Jump}: the first is due to the Korteweg stress and gives rise to a stress opposing the internal pressure
if the coefficient $\hat{\delta}$ of the Korteweg stress has a negative sign. This would mimic the effect of a positive tension.
The second is negative for a water-glycerin system and is proportional to the rate of change of viscosity with volume fraction.
While this term has the "wrong" sign for interfacial tension in the case of water and glycerine, it becomes positive
if the less dense fluid is more viscous than the denser one \cite{joseph_tension-joseph.pdf_1996}.

The continuum fluid dynamic approach predicts the possible existence of a positive tension that would regularize the problem of many
well known hydrodynamic instabilities in miscible fluids. However, this approach is affected by two limits:
1) the coefficients in the Korteweg stress are not known a priori and currently there are no means to calculate them starting from first principles;
2) the expression derived by Joseph, based on the solution of the continuum diffusion equation, is singular for $t=0$,
suggesting that such solution is unphysical for short times, precisely when Korteweg stresses are expected to contribute the most
to the interfacial dynamics before the interface smears out due to diffusion.
As we will see in the following section, both issues, are solved if miscible interfaces are considered in local equilibrium
so that standard statistical thermodynamics methods can be applied.

\subsection{Thermodynamic formulation: the case of binary mixtures}\label{SG-binary}
An alternative way to understand the existence of stresses at the boundary between miscible fluids is to take advantage of local equilibrium
in the framework of non-equilibrium thermodynamics. When a system globally out of equilibrium can be spatially and temporally divided into 'cells'
or 'micro-phases' of small (infinitesimal) size, in which classical equilibrium conditions are fulfilled to good approximation,
local equilibrium can be rigorously defined. These conditions do not hold, e.g., in very rarefied gases in which molecular collisions
are infrequent. By contrast they can be invoked for fluids. Once these 'cells' are defined, one admits that matter and energy
may pass freely between contiguous 'cells',
slowly enough to leave the 'cells' in their respective individual local thermodynamic equilibria with respect to intensive variables.

A typical example where local thermodynamic equilibrium exists is a glass of water that contains a melting ice cube. The temperature
can be defined at any point, but it is lower near the ice cube than far away from it. If the energy of the molecules
located near a given point could be observed, they will be distributed according to the Maxwell$-$Boltzmann (M$-$B) distribution.
Changing the observation region would lead to the same (M$-$B) distribution, albeit with a different temperature.
In synthesis, locally (and temporarily) the fluid molecules would in equilibrium.

In this and many other similar cases one can think of two 'relaxation times' separated by orders of magnitude \cite{Zubarev1974}.
The longer relaxation time is of the order of magnitude of the time taken by the macroscopic structure of the system to change.
The shorter one is of the order of magnitude of the time taken for a single 'cell' to reach local thermodynamic equilibrium.
If two time scales are well separated,
one may define a local free energy functional that plays a role analogous to equilibrium free energy for non-equilibrium
microstates \cite{jinwoo_local_2015}.

In this framework we consider a miscible interface, characterized by a gradient of concentration, for which we can define a free energy
of mixing \cite{kheniene_linear_2013,TruzzolilloPRX2016} $F$ as:
\begin{equation}\label{free_energy}
    F=U-TS=\int_V f_0(\varphi)+\frac{k_2(\varphi)}{2}(\nabla\varphi)^2+ O((\nabla\varphi)^4)
\end{equation}
where $U$, $T$, $S$ and $V$ represent respectively energy, entropy, temperature and volume of the micro-system and $\phi$ the volume fraction
of one the two type of molecules forming the two-fluid system.

It's worth noting that, for symmetry reasons, Eq. \ref{free_energy} must contain only even powers of the compositional gradient and no higher derivative terms: indeed, Cahn and Hilliard \cite{cahn58} argued that the derivative terms with odd powers should vanish.
Following Korteweg, one further assumes small compositional gradients and truncates the gradient expansion in Eq. \ref{free_energy} by neglecting
the terms of order higher than the square gradient term. In this case, $k_2(\varphi)$, often referred to as the capillary coefficient
for miscible fluids \cite{kheniene_linear_2013}, can be neglected everywhere except where there are large concentration gradients,
i.e. at the fluid boundaries. Together with the magnitude of the concentration gradient $\nabla \varphi$, $k_2$ univocally
determines the interfacial tension \cite{vanderwaals1894,Rowlinson82,Evans2009}.
Assuming that only the local part of the free-energy contributes to the effective interfacial tension $\Gamma_e$,
the latter can be calculated as

\begin{equation}\label{thermo_int}
    \Gamma_e=\frac{\partial F}{\partial A}=\int_{-\delta/2}^{\delta/2}k_2(\varphi)(\nabla\varphi)^2 dz
\end{equation}
where $z$ is the coordinate normal to the interface, $A$ is the interfacial area and $\delta$ is
the interface thickness, i.e. the region of the space where $|\nabla \varphi|>0$.

\subsection{On-lattice model for the calculation of the Korteweg coefficient}\label{lattice}
Equations \ref{free_energy} and \ref{thermo_int} are general and can, in principle, be used to calculate the interfacial tension
for two simple or complex fluids. Unfortunately the calculation of the square gradient coefficient $k_2(\varphi)$ is often not an easy task,
owing to the specific structure of the two fluids in contact. However, in some simplified cases, the calculation is possible and helps to understand the role played by parameters such as temperature, concentration and binding energies between the molecules.
One of these cases is a on-lattice binary mixture (Figure \ref{Lattice-Model}) which is a useful model for a molecular two-fluid system.

Following \cite{TruzzolilloPRX2016} the mixture is modeled by a lattice composed of cells of volume $\Omega$, occupied either by a particle of
type A, with average probability $\varphi$, or by a B particle, with average probability $1-\varphi$.
$\Gamma_e$ is calculating by computing the contribution to entropy and internal energy due to the presence of a region with a non zero gradient
of concentration. To that end one considers three lattice layers orthogonal to $z$ and labelled by the indexes $i-1$, $i$, and $i+1$,
as shown in Figure \ref{Lattice-Model}.
To take into account the various probabilities of AA, AB, and BB bonds within the three-layer region, it is convenient to indicate by $z'$
the number of neighbors of a site of the central layer that belongs to the adjacent layers $i \pm 1$, where $z'$ may in general be
different from the lattice coordination number $z$, so that the number of neighbors within the same layer is $z-2z'$.
Neglecting terms of order $O(|\nabla \varphi|^4)$ and after some algebra (see \cite{TruzzolilloPRX2016} for details)
the full expression of the square
gradient contribution to the interfacial tension reads:

\begin{figure}[htbp]
   \centering{\includegraphics[width=8cm]{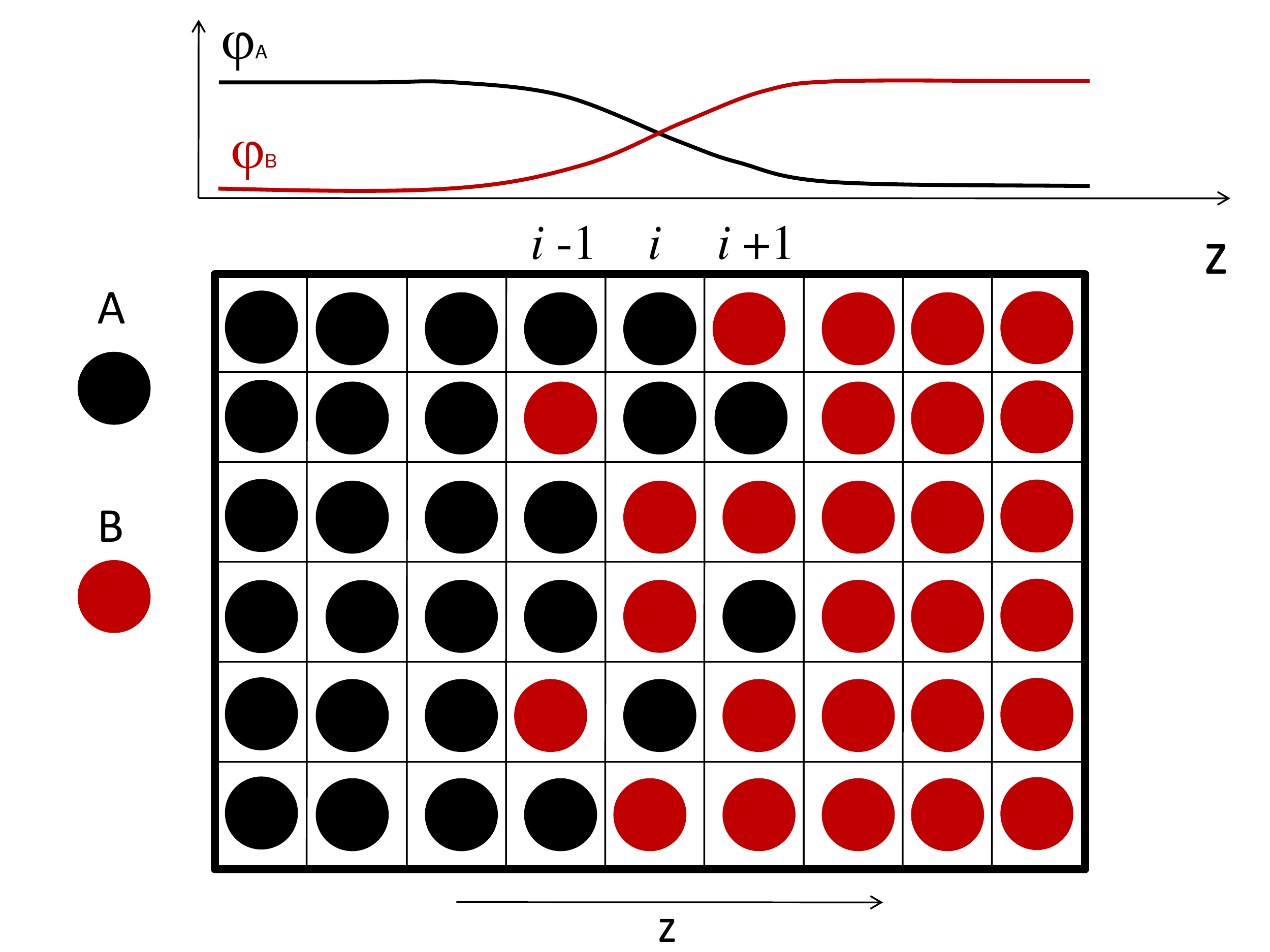}}\\
  \caption{Sketch of a binary mixture on lattice. Particles of type $A$ and type $B$, characterized by concentrations $\varphi_A$
  and $\varphi_B$ respectively
  are separated by a region of finite width where mixing occurs.}\label{Lattice-Model}
\end{figure}

\begin{equation}\label{Gammamf}
\Gamma_e=\frac{k_bTb^2}{\Omega\delta}\left\{\frac{z'}{z}\chi\frac{\varphi^2}{3}+\frac{2}{3}
\left[1+\frac{1-\varphi}{\varphi}\ln(1-\varphi)\right]\right\},
\end{equation}
where $\chi$ is the Flory-Huggins parameter of the mixture \cite{Flory1953}, $\Omega$ is the volume of each lattice site and $b$
the lattice constant. To the best of our knowledge, this is the only existing analytical expression of $\Gamma_e$ for simple mixtures.

\begin{figure}[htbp]
  % Requires\usepackage{graphicx}
\centering{\includegraphics[width=8cm]{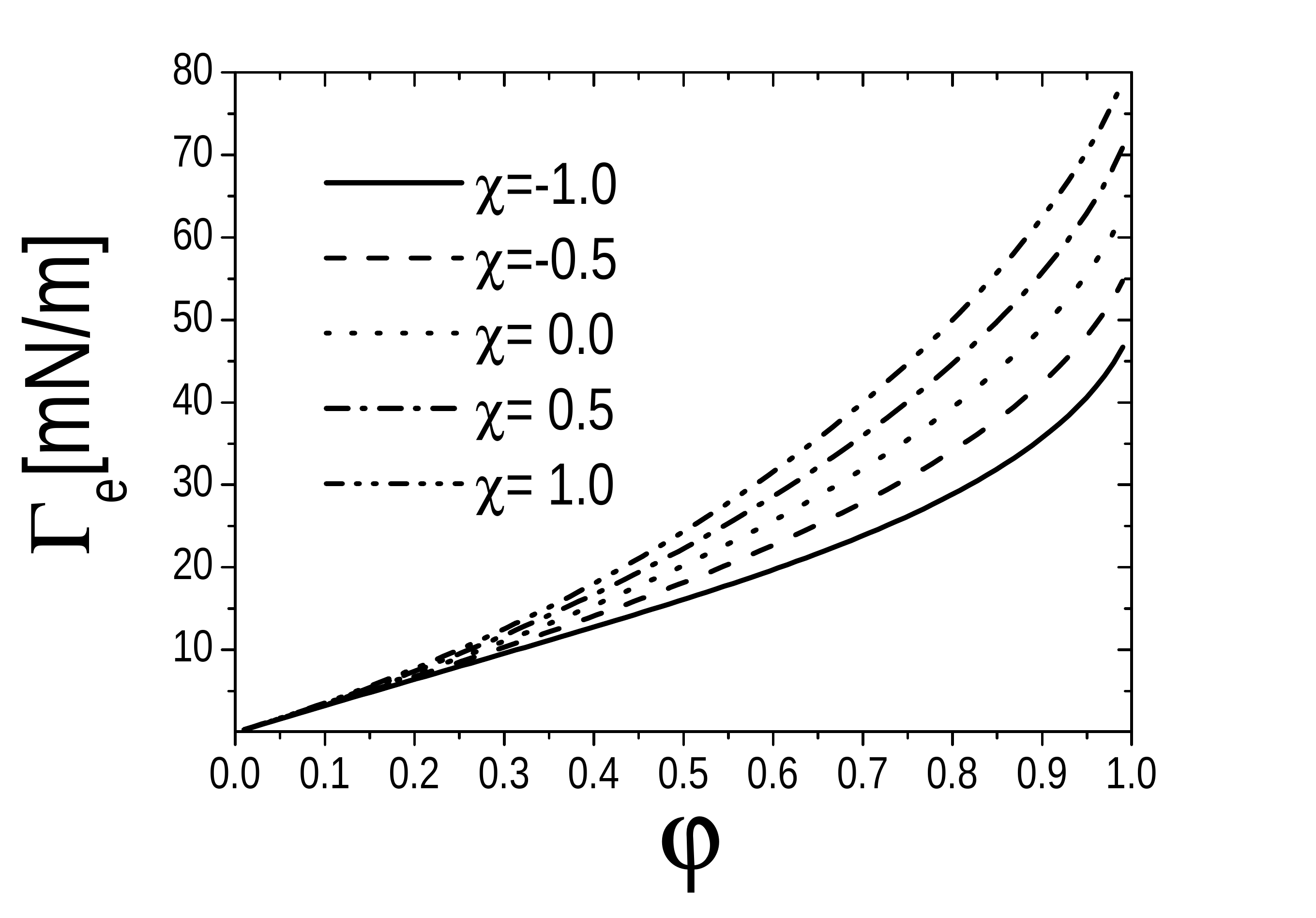}}\\
\caption{Effective interfacial tension $\Gamma_e(\varphi)$ calculated according to Eq. \ref{Gammamf} for
different values of Flory-Huggins parameter $\chi$ as shown in the label. Here $b=1.25$ ${\AA}$, $\Omega=\pi/6 b^3$,
$\delta=5b$ and $z'/z=1/2$.}\label{Gamma-molecular}
\end{figure}

We note that the square gradient approximation used to derive Eq. \ref{Gammamf} implies $\varphi\neq 0$,
$\varphi\neq 1$ and that the concentration gradient at the interface be small, i.e. that the effective surface tension be
dominated by the first term of Eq. \ref{free_energy}. Eq. \ref{Gammamf} suggests that

\begin{itemize}
\item $\Gamma_e$ is an increasing function of the bulk volume fraction the two species for values of the Flory-Huggins parameter typical
of real liquid mixtures (Fig. \ref{Gamma-molecular}).
\item The effective interfacial tension decreases as the interfacial thickness decreases. This is indeed what happens when time
passes and the mixture goes towards global equilibrium: the gradient of concentration and the stress anisotropy decreases at the boundary between
the fluids.
\item The higher the temperature, the larger the free energy cost to create a concentration gradient and thus $\Gamma_e$.
This can be understood by the fact that short
wavelength density fluctuations are suppressed upon increasing the temperature.
\item The higher the $\chi$-parameter, the larger the interfacial tension (see Fig. \ref{Gamma-molecular}). This occurs because $\chi$ tunes
the "affinity" between the two fluids: for high positive values of $\chi$,
the A-B contacts become more favorable
with respect to the A-A or B-B ones. Accordingly, segregation of the two species that must be associated with a concentration
gradient becomes more costly.
\item The effective tension can be of the same order of magnitude of those measured between immiscible fluids, i.e. tens of mN/m,
if the thickness of the interface is limited to few molecular diameters.
\end{itemize}

Although such features predicted by the lattice theory are all reasonable and physically consistent, they have not been tested experimentally
so far for molecular fluids, whose interfacial characterization is still a challenging task. Moreover it is worth emphasizing that Eq.
\ref{Gamma-molecular} gives an expression for $\Gamma_e$ based on the square gradient approximation. Consequently it is not expected to predict
quantitatively the magnitude of interfacial stresses for very sharp concentration gradients, for which other approaches are needed \cite{TruzzolilloPRX2016}.

\section{Conclusions}
We have given an overview of the most common hydrodynamic instabilities arising at the boundary between miscible fluids.
We have addressed a number
of questions concerning the mathematics and the physics of miscible interfaces, whose stability impacts the flow
of fluids in a wide variety of configurations, relevant to industrial applications as well as
to the deep understanding of natural phenomena occurring at vastly different scale.
In particular we have discussed how the lack of an equilibrium surface tension often rises the issue of the ill-posedness of the problem, as
defined in the work of Hadamard. Particular attention has been devoted to stratified and displacing miscible fluids whose stability,
together with the search for unambiguous
signatures of capillary forces, currently represents an important part of the experimental research in our group.
It is therefore no accident that we have dedicated one full section to Korteweg stresses, whose importance for
the stability and the regularization of the mathematical formulation of the behaviour of miscible interfaces
is a lively debated issue.
Hydrodynamic instabilities are an extremely wide research area: the topics discussed here are of course just a small selection, driven
by our own curiosity and research activity.
Nevertheless, it is our hope that this review will be helpful to scientists working on multi-fluid flows and contribute to attract
new researchers to this fascinating field.

\section{Acknowledgments}
D.T. is very grateful to Dr I. Bischofberger and Prof. R. Govindarajan for useful discussions.\\

\bibliographystyle{iopart-num}
%\bibliography{biblio-review,manuscript,manuscript-sup}
\providecommand{\newblock}{}

\end{document}